\documentclass[%
 aip,
 amsmath,amssymb,
 reprint,%
 twocolumn,%
]{revtex4-1}

\usepackage{graphicx}
\usepackage{dcolumn}
\usepackage{bm}

\usepackage[utf8]{inputenc}
\usepackage[T1]{fontenc}
\usepackage{mathptmx}
\usepackage{color}
\begin{document}


\author{A. Eugene DePrince III}
\affiliation{Department of Chemistry and Biochemistry, Florida State University, Tallahassee, FL 32306-4390}
\email{adeprince@fsu.edu}

\title{Cavity-modulated ionization potentials and electron affinities from quantum electrodynamics coupled-cluster theory}

	
\begin{abstract}
	Quantum electrodynamics coupled-cluster (QED-CC) theory is used to model vacuum-field-induced changes to ground-state properties of a series of sodium halide compounds (NaX, X = F, Cl, Br, I) strongly coupled to an optical cavity. Ionization potentials (IPs) and electron affinities (EAs) are presented, and it is demonstrated that EAs are easily modulated by cavity interactions, while IPs for these compounds are far less sensitive to the presence of the cavity. EAs predicted by QED-CC can be reduced by as much 0.22 eV (or $\approx 50\%)$ when considering experimentally-accessible coupling parameters.
\end{abstract}
	
\maketitle
	
\section{Introduction}

Strong interactions between nanoconfined photons and molecular systems\cite{Nori19_19,Narang18_1479,Barnes14_013901} can lead to the creation of hybrid light--matter states known as polaritons that may display remarkably different chemical and physical properties than their parent components.\cite{Whittaker98_6697,Mugnier04_036404,Ebbesen12_1592,Smith14_5561,Ebbesen15_1123,Baumberg16_127,Ebbesen16_2403,Nitzan17_443003,Ebbesen17_9034,Bellessa19_173902,Forrest20_371} The technological and chemical applications of these strongly-coupled light--matter states are wide-ranging and include cavity-based catalysis\cite{George19_10635,Feist19_8698}, polariton lasing\cite{Forrest20_371}, long-range optical energy propagation,\cite{Basu20_5043} and plasmon-based photostabilization\cite{Shegai18_eaas9552}. Aside from the creation of polariton states, embedding a molecule within an optical cavity can also lead to interactions between the molecule and vacuum field fluctuations that alter ground-state properties such as electron density distributions in small molecules\cite{Rubio18_992,Koch20_041043} or transition temperatures in spin-crossover complexes.\cite{Feist19_8698}

Theoretical modeling is a crucial tool in the design of cavity-based technologies with novel or enhanced optical or catalytic properties.
Historically, theoretical descriptions of strong light--matter interactions have been built upon simple model Hamiltonians which describe interactions between few-level quantum emitters and dipolar photon modes. Such models are powerful and offer essential insight into qualitative changes to optical properties of molecular systems strongly coupled to nanoconfined photons,\cite{Carusotto06_033811,Genes19_203602,Nori19_19,Gray13_075411,Huo19_5519,Huo20_6321,Foley20_9063} but they are hardly suitable for discerning subtle changes to electronic structure, particularly of the ground-state. Quantitative predictions of chemical reactivity or orbital-specific quantities ({\em e.g.}, ionization potentials or electron affinities) within optical cavities or other nanoconfined environments necessitate an {\em ab inito} approach to light--matter interactions.

The most popular and well-developed polaritonic quantum chemical model is the quantum electrodynamics density functional theory (QEDFT) approach,\cite{Bauer11_042107,Tokatly13_233001,Rubio14_012508,Rubio15_093001,Rubio18_992,Appel19_225,Narang20_094116} which as been applied successfully to a variety of interesting problems involving strong light-matter interactions. However, QEDFT is not a panacea for the polaritonic problem, for two reasons. First, the exact form of the polaritonic exchange-correlation functional in QEDFT is unknown, although promising recent work\cite{Rubio15_093001,Rubio18_992} has generalized the optimized effective potential approach in an attempt to remedy this shortcoming. Second, being an extension of DFT for the electronic case, QEDFT is susceptible to all of the well-known deficiencies exhibited by DFT\cite{Yang08_792} that potentially limit its applicability to arbitrary polaritonic problems. As such, it becomes desirable to work within a systematically-improvable many-body framework, such as that afforded by coupled-cluster (CC) theory.\cite{Cizek66_4256,Paldus71_359,Bartlett09_book,Musial07_291}  
Polaritonic CC\cite{Manby20_023262} and quantum electrodynamics CC (QED-CC)\cite{Koch20_041043} are similar extensions of standard CC theory for the electronic problem {\color{black} that incorporate both photon and coupled photon-electron transitions}. Both formalisms retain the many nice properties of CC theory, including its systematic improvability and size-extensivity {\color{black} with respect to the electronic space}; the principal difference between them is the precise {\color{black} treatment of the photon degrees of freedom}. {\color{black}Polaritonic CC can be thought of as a mixture of CC for electrons and configuration interaction for photons; as a result, polaritonic CC will not be size-extensive in the photon space as more modes are considered. QED-CC, on the other hand, is a fully size-extensive approach. As a more minor distinction, polaritonic CC has only been applied with model Hamiltonians, whereas the QED-CC model of Ref.~\citenum{Koch20_041043} considered a fully  {\em ab initio} electronic Hamiltonian.}

In this work, we apply the QED-CC formalism of Ref.~\citenum{Koch20_041043} to the evaluation of ground-state properties in molecules strongly-coupled to optical cavities. Specifically, we consider whether cavity interactions can effect meaningful changes to ionization potentials (IPs) or electron affinities (EAs) of the sodium halide series (NaX, X = F, Cl, Br, I), and we find that, in the case of EAs, changes as large as 0.22 eV can be realized, using experimentally accessible cavity parameters. Changes in EAs predicted by quantum electrodynamics Hartree-Fock (QED-HF) theory are even larger, and, in the case of NaF, cavity interactions can even change the sign of the QED-HF-derived EAs. On the other hand, we find that IPs for these molecules are significantly less sensitive to cavity interactions.

This paper is organized as follows. Section \ref{SEC:THEORY} presents an overview of QED-CC theory restricted to coupled single and double electronic excitations and single photon excitations (termed QED-CCSD-1 in Ref.~\citenum{Koch20_041043}), and Sec.~\ref{SEC:COMPUTATIONAL_DETAILS} outlines the details of our computations. 
In Sec.~\ref{SEC:RESULTS}, we apply QED-CCSD-1 to the evaluation of IPs and EAs in the sodium halide series, and some concluding remarks can be found in Sec.~\ref{SEC:CONCLUSIONS}.

\section{Theory}

\label{SEC:THEORY}

In this Section, we present an overview of QED-CC theory, limited to coupled single and double electronic excitations and single photon excitations (QED-CCSD-1), following the formalism outlined in Ref.~\citenum{Koch20_041043}. We restrict our considerations to interactions between a molecule and a single dipolar cavity mode, for which we 
define the Pauli--Fierz Hamiltonian\cite{Spohn04_book,Rubio18_0118} (in atomic units), within the dipole approximation and in the length gauge:
\begin{equation}
    \label{EQN:PFH}
    \hat{H} = \hat{H}_{\rm e} + \omega_{\rm cav} \hat{b}^\dagger \hat{b} - \bigg ( \frac{\omega_{\rm cav}}{2} \bigg )^{1/2} ({\bm \lambda} \cdot {\bm \mu})(\hat{b}^\dagger +\hat{b}) + \frac{1}{2} ({\bm \lambda} \cdot {\bm \mu})^2
\end{equation}
Here, $\hat{H}_{\rm e}$ represents the usual Hamiltonian for the electronic component of the system, 
\begin{equation}
    \hat{H}_{\rm e} = \sum_{pq} h_{pq} \hat{a}_p^\dagger \hat{a}_q + \frac{1}{2}\sum_{pqrs} g_{pqrs} \hat{a}^\dagger_p \hat{a}^\dagger_r \hat{a}_s \hat{a}_q,
\end{equation}
where $h_{pq}$ is an element of the matrix representation of the core Hamiltonian, $g_{pqrs}$ is a two-electron repulsion integral, $\hat{a}^\dagger$ and $\hat{a}$ represent fermionic creation and annihilation operators, respectively, and the labels $p$, $q$, $r$, and $s$ refer to spin orbitals. The second term in Eq.~\ref{EQN:PFH} represents the effective Hamiltonian for the cavity modes, which are harmonic oscillators with fundamental frequencies $\omega_{\rm cav}$ and coupling vectors ${\bm \lambda}$. The operator $\hat{b}^\dagger$ ($\hat{b}$) is a bosonic creation (annihilation) operator that creates (destroys) one photon in the cavity mode. The third term in Eq.~\ref{EQN:PFH} represents dipolar coupling between the electron and photon degrees of freedom, and the last term is the molecular dipole self-energy. The symbol ${\bm \mu}={\bm \mu}_{\rm e} + {\bm \mu}_{\rm nu}$ represents the total (electronic + nuclear) dipole operator for the molecular component of the system. We consider only fixed nuclei ({\em i.e.}, the cavity Born-Oppenheimer approximation\cite{Rubio17_1616}), so we have neglected the nuclear repulsion and nuclear kinetic energy terms in the Hamiltonian given in Eq.~\ref{EQN:PFH}.

The QED-HF theory assumes only mean-field coupling between the photon and electron degrees of freedom, and the ground-state wave function is thus a direct-product of a Slater determinant and a (zero-photon) photon-number state
\begin{equation}
\label{EQN:QEDHF_REFERENCE}
    |\Psi_{\rm 0,0}\rangle = |\Psi_{\rm e,0}\rangle \otimes|\Psi_{\rm ph,0}\rangle,
\end{equation}
The electronic orbitals that comprise the Slater determinant $|\Psi_{\rm e,0}\rangle$ can be determined via a modified Hartree-Fock procedure that accounts for the presence of the cavity (see Ref.~\citenum{DePrince15_214104}, for example), and $|\Psi_{\rm ph,0}^A\rangle$ is the lowest-energy eigenfunction of the Hamiltonian
\begin{equation}
\label{EQN:H_PHOTON}
    \hat{H}_{\rm ph+e} = \omega_{\rm cav} \hat{b}^\dagger \hat{b} -\bigg ( \frac{\omega_{\rm cav}}{2} \bigg )^{1/2} ( {\bm \lambda} \cdot \langle {\bm \mu}\rangle)(\hat{b}^\dagger +\hat{b})
\end{equation}
which can be expanded in a basis of photon-number states. The expectation value of the dipole operator is taken with respect to $|\Psi_{\rm e,0}\rangle$. For convenience, we choose to follow Ref.~\citenum{Koch20_041043} and transform $\hat{H}$ and $\hat{H}_{\rm ph+e}$ to the coherent-state basis, which, in practice, simply results in a shift of the dipole operator by $-\langle {\bm \mu}\rangle$. Consequently, the second term in Eq.~\ref{EQN:H_PHOTON} vanishes, and the Pauli--Fierz Hamiltonian becomes
\begin{eqnarray}
    \label{EQN:PFH_COHERENT}
    \hat{H} &=& \hat{H}_{\rm e} + \omega_{\rm cav} \hat{b}^\dagger \hat{b} - \bigg ( \frac{\omega_{\rm cav}}{2} \bigg )^{1/2} ({\bm \lambda} \cdot [{\bm \mu} - \langle {\bm \mu }\rangle] )(\hat{b}^\dagger +\hat{b}) \nonumber \\
    &+& \frac{1}{2} ({\bm \lambda} \cdot [{\bm \mu} - \langle {\bm \mu }\rangle] )^2
\end{eqnarray}
We solve the QED-HF and subsequent QED-CC problems in this coherent-state basis, using this transformed Hamiltonian.

The ground-state QED-CC wave function is 
\begin{equation}
    |\Psi_{\rm CC}\rangle = e^{\hat{T}}|\Psi_{\rm 0,0}\rangle,
\end{equation}
with the cluster operator defined as
\begin{equation}
    \hat{T} = \hat{T}_{\rm e} + \hat{T}_{\rm ph} + \hat{T}_{\rm e,ph}
\end{equation}
Here, $\hat{T}_{\rm e}$ is the usual electronic cluster operator, and $\hat{T}_{\rm ph}$ and $\hat{T}_{\rm ph,e}$ are photon and simultaneous photon/electron cluster operators, respectively. At the QED-CCSD-1 level of theory\cite{Koch20_041043}, we have
\begin{equation}
    \hat{T}_{\rm e} = \sum_{ia} t_i^a \hat{a}^\dagger_a \hat{a}_i  + \frac{1}{4} \sum_{ijab} t_{ij}^{ab} \hat{a}^\dagger_a \hat{a}^\dagger_b \hat{a}_j \hat{a}_i, 
\end{equation}
~
\begin{equation}
    \hat{T}_{\rm ph} = u \hat{b}^\dagger
\end{equation}
and
\begin{equation}
    \hat{T}_{\rm e,ph} = \sum_{ia} u_i^a \hat{a}^\dagger_a \hat{a}_i \hat{b}^\dagger + \frac{1}{4} \sum_{ijab} u_{ij}^{ab} \hat{a}^\dagger_a \hat{a}^\dagger_b \hat{a}_j \hat{a}_i \hat{b}^\dagger
\end{equation}
where $t_i^a$, $t_{ij}^{ab}$, $u$, $u_i^a$, and $u_{ij}^{ab}$ are cluster amplitudes. The labels $i$ and $j$ ($a$ and $b$) refer to electronic orbitals that are occupied (unoccupied) in the QED-HF reference function. The energy and cluster amplitudes are determined in the usual way, by solving coupled nonlinear equations generated via projection of the Schr\"{o}dinger equation
\begin{equation}
    \hat{H} e^{\hat{T}}|\Psi_{\rm 0,0}\rangle = E e^{\hat{T}}|\Psi_{\rm 0,0}\rangle
\end{equation}
onto the relevant spaces:
\begin{eqnarray}
\langle \Psi_{m,n} | e^{-\hat{T}} \hat{H} e^{\hat{T}}|\Psi_{\rm 0,0}\rangle  = E \delta_{m,0} \delta_{n,0} \\
\end{eqnarray}
Here, $m$ refers to an electronic configuration (the reference configuration or singly / doubly substituted configurations), and $n$ refers to a photon number state (0 or 1). More explicit equations, using a slightly different notation, can be found in the Appendix of Ref.~\citenum{Koch20_041043}.

\section{Computational Details}

\label{SEC:COMPUTATIONAL_DETAILS}

The {\color{black} unrestricted QED-HF and} QED-CCSD-1 method{\color{black}s were} implemented {\color{black}in C++} in \texttt{hilbert},\cite{hilbert} {\color{black} which is} a {\color{black}free and open-source} plugin to the \textsc{Psi4} electronic structure package.\cite{Sherrill20_184108} All calculations employed the def2-TZVPPD basis set. The electron repulsion integral (ERI) tensor was approximated using the density fitting {\color{black}(DF)} approach,\cite{Whitten73_4496,Sabin79_3396} and the auxiliary basis sets used in the QED-HF and QED-CC portions of the calculations were the def2-universal-JKFIT and def2-TZVPPD-RIFIT basis sets, respectively. Following Ref.~\citenum{Koch20_041043}, our QED-CCSD-1 formalism employs a $t_1$-dressed Hamiltonian,\cite{Helgaker94_233} and the $t_1$-transformation uses the def2-TZVPPD-RIFIT basis set for all terms; this strategy differs slightly from that {\color{black}described in} Ref.~\citenum{Sherrill13_2687}, which used JK- and RI-type basis sets in the $t_1$-transformation of the Fock matrix and all other terms, respectively. {\color{black}The combined use of DF and a $t_1$-dressed Hamiltonian allows for the design of an efficient implementation, in terms of both floating-point and memory costs. For example, DF reduces the floating-point costs associated with the $t_1$-transformation of the Hamiltonian from fifth-power to fourth-power with respect to system size, and the remaining working equations of QED-CCSD-1 can be structured so as to avoid the explicit storage of blocks of the ERI tensor involving more than two virtual labels. As a result, the storage requirements of our algorithm scale as $\mathcal{O}(o^2v^2)$+$\mathcal{O}(N_{\rm aux}(o+v)^2)$, where $o$ and $v$ represent the number of occupied and virtual spin orbitals, respectively, and $N_{\rm aux}$ represents the dimension of the auxiliary basis set} 

Geometries for the isolated neutral compounds were optimized at the density functional theory (DFT) level of theory, using the B3LYP exchange-correlation functional and the def2-TZVPPD basis set. The DFT calculations made use of the density fitting approximation to the ERI tensor, with the def2-universal-JKFIT auxiliary basis set. Lastly, note that calculations involving iodine employed the def2-TZVPPD effective core potential.

\section{Results and Discussion}

\label{SEC:RESULTS}

We now investigate the degree to which cavity interactions can modulate ground-state properties in a series of sodium halide compounds (NaX, X = F, Cl, Br, I). Specifically, we consider changes to ionization potentials, electron affinities, and the electronic densities upon the introduction of non-zero cavity coupling parameters. In all calculations, we consider a cavity mode with frequency $\omega_{\rm cav}=2.0$ eV that is polarized in the $z$ direction [{\em i.e.}, {$\bm \lambda$} = (0, 0, $\lambda)$]. Tables \ref{TAB:IP} and \ref{TAB:EA}, respectively, present IPs and EAs for the sodium halide series (with the compounds oriented in the $z$-direction, parallel to the polarization of the cavity mode) for coupling parameters in the range $\lambda$ = 0.0 to $\lambda$ = 0.05. $\Delta$QED-CC and $\Delta$QED-HF IPs and EAs are computed via energy differences among the charged and neutral species, {\color{black}{\em i.e.},
\begin{eqnarray}
{\rm EA} = E(N) - E(N-1) \\
{\rm IP} = E(N+1) - E(N),
\end{eqnarray}
where $E(N)$, $E(N-1)$, and $E(N+1)$ represent the energy of the neutral, anion, and cation species, respectively, evaluated at the relevant level of theory.
}These IPs and EAs are vertical; the optimized (B3LYP) geometries for the neutral species, in the absence of the cavity, are used for neutral, cation, and anion calculations.

\begin{table}[!htpb]
    \caption{$\Delta$QED-HF and $\Delta$QED-CC IPs (eV) for sodium halide compounds coupled to a cavity with a fundamental frequency $\omega_{\rm cav}$=2.0 eV, polarized along the molecular axis.}
    \label{TAB:IP}
    \begin{center}
        \begin{tabular}{cccccccccc}

            \hline\hline
        
  	        & \multicolumn{4}{c}{$\Delta$QED-CC} &~~& \multicolumn{4}{c}{$\Delta$QED-HF} \\	
			\cline{2-5} \cline{7-10}					
      $\lambda$    	&NaF	&NaCl	&NaBr	&NaI	&&NaF	&NaCl	&NaBr	&NaI    \\
      	\hline
0.00	&9.96	&9.00	&8.55	&7.98	&&8.10	&7.99	&7.68	&7.18 \\
0.01	&9.96	&9.00	&8.55	&7.97	&&8.10	&7.98	&7.68	&7.17 \\
0.02	&9.95	&8.99	&8.54	&7.97	&&8.09	&7.97	&7.67	&7.16 \\
0.03	&9.95	&8.98	&8.53	&7.95	&&8.08	&7.96	&7.64	&7.13 \\
0.04	&9.93	&8.96	&8.51	&7.93	&&8.07	&7.93	&7.62	&7.09 \\
0.05	&9.92	&8.94	&8.49	&7.90	&&8.05	&7.90	&7.58	&7.05 \\
            \hline\hline
        
        \end{tabular}
    \end{center}

\end{table}

\begin{table}[!htpb]
    \caption{$\Delta$QED-HF and $\Delta$QED-CC EAs (eV) for sodium halide compounds coupled to an optical cavity with a fundamental frequency $\omega_{\rm cav}$=2.0 eV, polarized along the molecular axis.}
    \label{TAB:EA}
    \begin{center}
        \begin{tabular}{cccccccccc}

            \hline\hline
        
  	        & \multicolumn{4}{c}{$\Delta$QED-CC} &~~& \multicolumn{4}{c}{$\Delta$QED-HF} \\	
			\cline{2-5} \cline{7-10}					
      $\lambda$    	&NaF	&NaCl	&NaBr	&NaI	&&NaF	&NaCl	&NaBr	&NaI    \\
      	\hline
0.00	&0.43	&0.64	&0.70	&0.78	&&0.37	&0.56	&0.62	&0.68 \\
0.01	&0.42	&0.64	&0.70	&0.77	&&0.36	&0.55	&0.61	&0.67 \\
0.02	&0.40	&0.62	&0.68	&0.75	&&0.33	&0.52	&0.58	&0.64 \\
0.03	&0.36	&0.59	&0.65	&0.72	&&0.28	&0.47	&0.53	&0.59 \\
0.04	&0.32	&0.54	&0.60	&0.68	&&0.21	&0.41	&0.47	&0.54 \\
0.05	&0.26	&0.49	&0.55	&0.63	&&0.13	&0.34	&0.40	&0.47 \\
            \hline\hline
        
        \end{tabular}
    \end{center}

\end{table}

From Table \ref{TAB:IP}, it is clear that IPs in these compounds are not particularly sensitive to coupling to the vacuum field, at either level of theory. The small change that we do observe is a universal decrease in the IPs with increasing $\lambda$. The largest change predicted by QED-HF occurs for NaI with $\lambda = 0.05$, where the IP is 0.13 eV less than QED-HF predicts for the isolated compound. The largest change predicted by QED-CCSD-1 also occurs for NaI, but this change is smaller (only 0.08 eV less than the QED-CCSD-1 IP predicted for the isolated compound). 

Seeing as how IPs in these compounds are {\color{black}somewhat} insensitive to cavity interactions, we focus for the remainder of this work on the case of EAs, which appear to be much more responsive to vacuum field fluctuations. As shown in Table \ref{TAB:EA}, EAs are found to universally decrease with an increase in the coupling strength, for all compounds and at both levels of theory. The largest changes are observed for NaF, where, at the QED-CCSD-1 level of theory, the EA is predicted to decrease by 0.17 eV. QED-HF predicts even larger changes; decreases of more than 0.2 eV are predicted for all compounds, with the largest change (-0.24 eV) occuring for NaF.

\begin{figure*}[!htpb]
    \centering
    \includegraphics{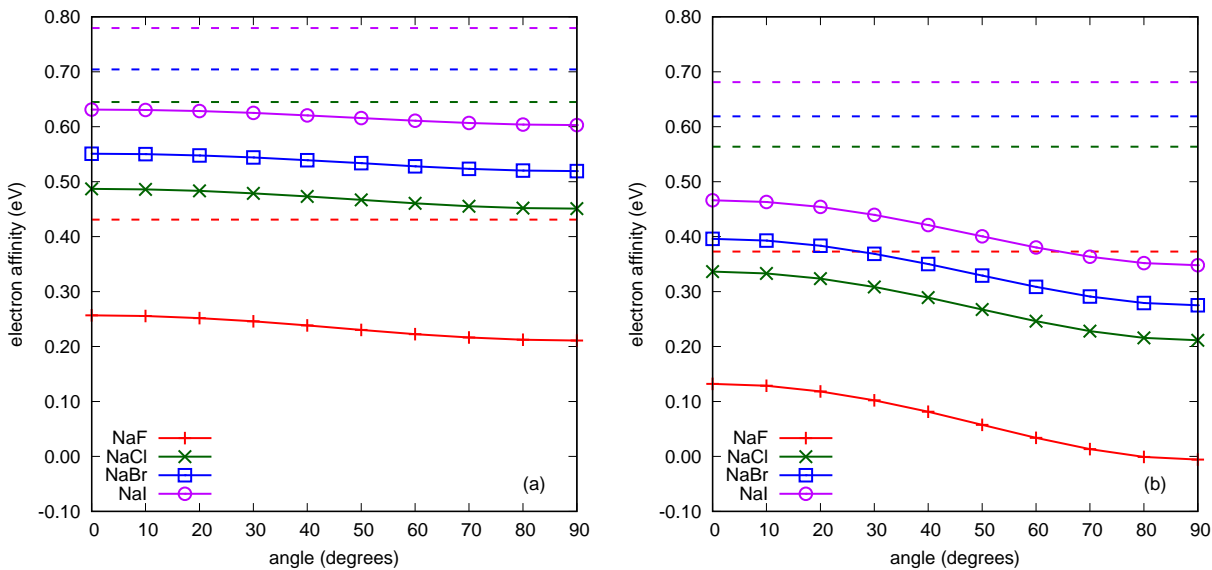}
    \caption{Electron affinities (eV) for sodium halide compounds as a function of the relative orientation of the molecular axis and the cavity mode, as predicted by QED-CCSD-1 (a) and QED-HF (b). Dashed lines represent the EA values computed for each compound in the absence of the cavity.}
    \label{FIG:ROTATE}
\end{figure*}

It is well known that the properties of strongly-coupled light-matter systems are highly dependent upon the geometry of the system ({\em e.g.}, the relative orientation of transition dipoles and cavity modes). Current cavity fabrication techniques do not allow for precise control over molecular orientation, so it is important to consider non-idealized geometries when predicting cavity-induced changes to molecular properties. Accordingly, in Fig.~\ref{FIG:ROTATE} we present the EAs for the sodium halide compounds as a function of the angle between the molecular axis and the polarization of the cavity mode (0$^\circ$ = aligned), with a cavity coupling strength $\lambda=0.05$. Dashed lines represent the EA values computed for each compound in the absence of the cavity. Interestingly, the maximum deviations between the EA of the cavity-embedded and isolated species are not realized when the molecular axis (and thus the permanent dipole moment) is aligned with the cavity mode; for all compounds, at both the QED-CCSD-1 and QED-HF levels of theory, the minimum EA observed occurs at an angle of 90$^{\circ}$. At the QED-CCSD-1 level of theory, the minimum values in the EAs for NaF, NaCl, NaBr, and NaI are 0.21 eV, 0.45 eV, 0.52 eV, and 0.60 eV, respectively. Recall that the EAs for the isolated compounds are 0.43 eV, 0.64 eV, 0.70 eV, and 0.78 eV. Hence, the largest change in the EA that can be realized at the QED-CCSD-1 level of theory is a decrease of 0.22 eV, in the case of NaF, which amounts to a roughly 50\% reduction in the isolated-compound EA. EAs predicted by QED-HF are even more sensitive to molecular orientation. At this level of theory, the minimum values in the EAs for NaF, NaCl, NaBr, and NaI are -0.01 eV, 0.21 eV, 0.28 eV, and 0.35 eV, respectively. The largest change relative to the isolated-compound limit is found to be that of NaF, where the EA actually becomes {\em negative} at an angle of 90$^{\circ}$; this result corresponds to a decrease in the EA by 0.38 eV relative to the case of the isolated compound. We note that we performed a similar analysis of the angle dependence of the IPs of these compounds, which, unsurprisingly, were found to be quite insensitive to the molecular orientation. At $\lambda = 0.05$, IPs derived from QED-CCSD-1 and QED-HF were found to change by, at most, only 0.01 eV and 0.02 eV, respectively, over the full range of compounds and orientations explored.

\begin{figure}[!htpb]
    \centering
    \includegraphics{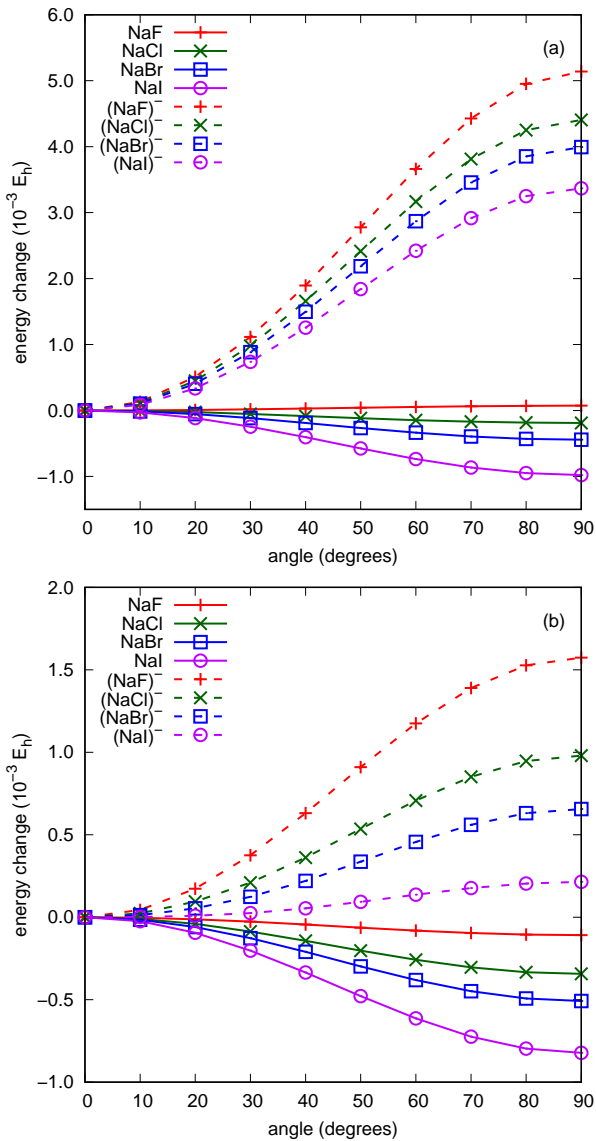}
    \caption{The change in energy (10$^{-3}$ E$_{\rm h}$) for neutral and negatively-charged sodium halides, as described by (a) QED-HF and (b) QED-CCSD-1 (b).}
    \label{FIG:ROTATE_ENERGY_CHANGE}
\end{figure}

It is interesting to consider what factors contribute to the changes in EAs ({\em i.e.}, stabilization / destabilization of the neutral / ionic species). Figure \ref{FIG:ROTATE_ENERGY_CHANGE} depicts changes in the total energies for NaX and (NaX)$^{-}$ computed at the QED-HF [panel (a)] and QED-CCSD-1 [panel (b)] levels of theory, relative to the relevant value computed when the compounds are aligned with the cavity mode (at 0$^\circ$). First, we consider the mean-field description of these species illustrated in panel (a). We find that QED-HF predict a stabilization of the neutral compounds (solid lines) as their alignment is brought from parallel to the cavity mode polarization to perpendicular to it, with the exception of NaF, for which there is a slight increase in the energy. On the other hand, the anion species (dashed lines) are universally destabilized as they are oriented perpendicularly to the cavity mode, and the degree of destabilization of (NaX)$^{-}$ is much greater than the degree of stabilization of NaX. These effects are synergistic and both contribute to the decrease in the EA, but it appears that, at the QED-HF level of theory, the decrease in the EAs observed when the compounds are oriented perpendicularly to the cavity mode polarization is driven mainly by the destabilization of the anionic species. The interplay between stabilization and destabilization are somewhat more subtle at the QED-CCSD-1 level of theory. As in the case of the mean field calculations, we observe general and synergistic stabilization/destabilization of the neutral/anion species as the compounds are brought out of alignment with the cavity mode polarization. However, the correlated calculations predict that, for a given species, the magnitudes of these changes are similar. As a result, we find that the decrease in the EA for NaF, NaCl and NaBr is mainly due to the destabilization of the anion species, while the stabilization of the neutral compound is the principal driver in the case of NaI.


\begin{figure*}[!htpb]
    \centering
    \includegraphics{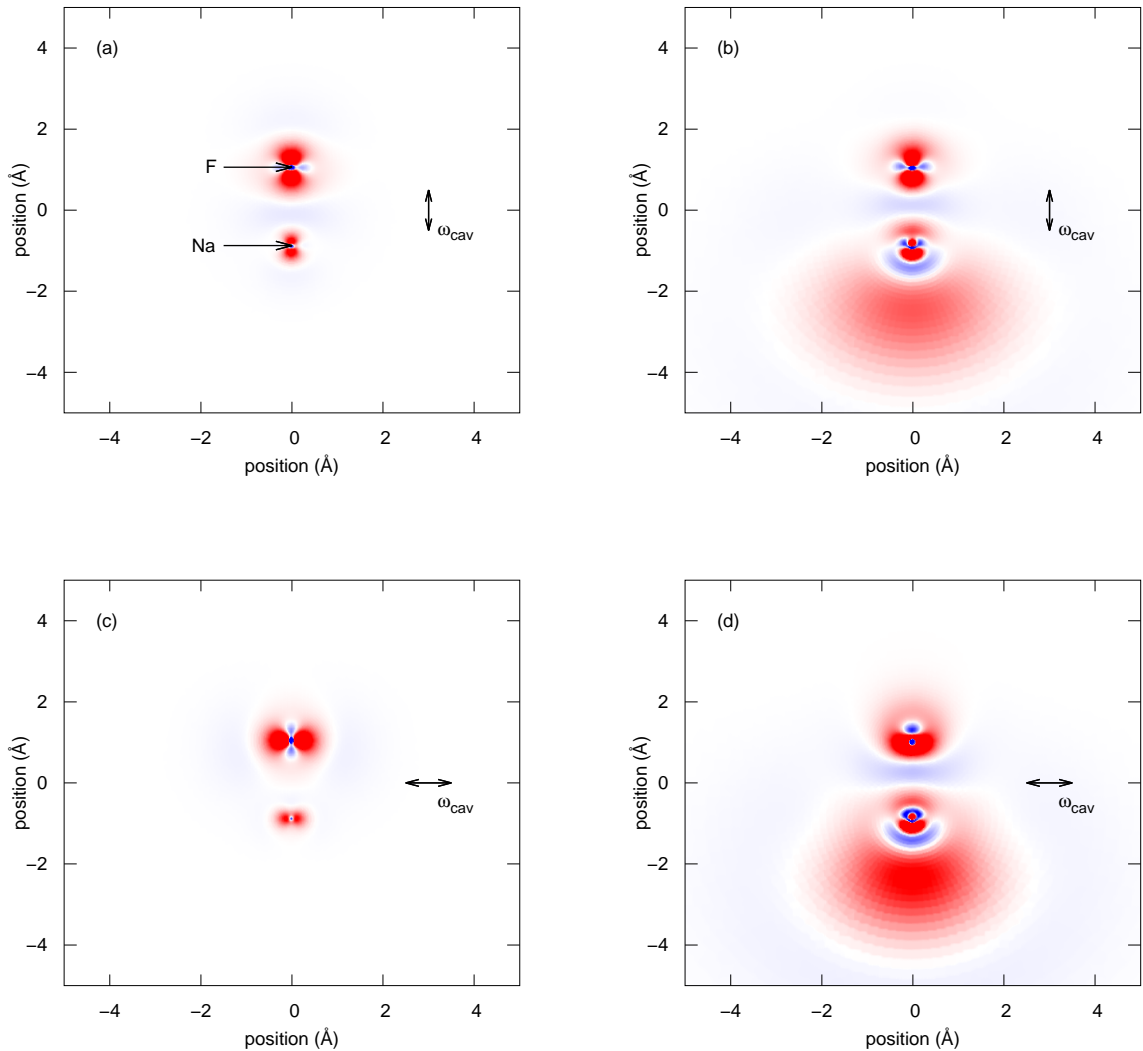}
    \caption{Density differences for isolated and cavity-embedded NaF and (NaF)$^{-}$. We consider 
    (a) $\rho^{\|}_{0} - \rho_{0} $,
    (b) $\rho^{\|}_{-1} - \rho_{-1} $,
    (c) $\rho^{\perp}_{0} - \rho_{0} $, and
    (d) $\rho^{\perp}_{-1} - \rho_{-1} $.}
    
    \label{FIG:DENSITY}
\end{figure*}

Lastly, we attempt to gain some insight into the destabilization of the sodium halide anion species relative to the neutral compounds through an analysis of cavity-induced changes to the density of the compound for which we observe the largest change in EA:  NaF as described by QED-HF theory. Figure \ref{FIG:DENSITY} provides slices of various difference densities involving isolated neutral NaF ($\rho_0$), isolated (NaF)$^{-}$ ($\rho_{-1}$), as well as these species embedded within a cavity with a cavity mode polarized along the molecular axis ($\rho^{\|}_0$ and $\rho^{\|}_{-1}$) or perpendicular to it ($\rho^{\perp}_0$ and $\rho^{\perp}_{-1}$). The positions of fluorine and sodium centers are indicated with arrows in panel (a). The color scheme indicates an increase in electron density with red and a decrease in electron density in blue, and all panels use the same color scale.

First, we consider the case of neutral NaF and the change in the density induced by a cavity mode polarized along the molecular axis [panel (a) of Fig.~\ref{FIG:DENSITY}]. The primary changes are the reduction of electron density at both the fluorine and sodium atom centers, along with increases in density in lobes polarized along the cavity mode axis. At fluorine, small decreases in electron density can also be observed in a lobes oriented perpendicular to the cavity mode polarization. When the compound is oriented perpendicular to the cavity mode polarization [panel (c)], we observe similar changes. The cavity induces in a reduction of electron density at the atom centers, and that density accumulates primarily in lobes oriented along the cavity mode polarization axis. Again, at fluorine, lobes indicating the loss of electron density are orientated perpendicular to the cavity mode. Recall from Fig.~\ref{FIG:ROTATE_ENERGY_CHANGE} that QED-HF predicts only a minor difference in energy between NaF orientations along the cavity mode axis or perpendicular to it. The similarity of the energies at these geometries is consistent with the similar electron displacements observed here.

Panel (b) of Fig.~\ref{FIG:DENSITY} depicts the cavity-induced changes to the electronic density of (NaF)$^{-}$ when the cavity mode is polarized along the molecular axis. Changes at the fluorine center are similar to those observed for neutral NaF in panel (a), and much more complex structure emerges at the sodium center. At sodium, we note that the lobes of increased electron density discussed in panel (a) persist, but they are deformed, with more density shifted toward the exterior of the compound. Additionally, some nodal structure emerges, with alternating electron rich and deficient regions that extend much farther than the deformations exhibited in the vicinity of sodium in the neutral species. When the cavity mode is polarized perpendicular to the molecular axis [panel (d) of Fig.~\ref{FIG:DENSITY}], the cavity-induced changes to the anion density around sodium are similar to those observed in panel (b). This result contrasts with the case of the neutral NaF where, in panels (a) and (c), we observe that the lobes of electron-rich regions are oriented along cavity mode polarization axis. Moreover, the density deformations in the vicinity of fluorine are distinct from any we have discussed to this point. First, the lobe structure present in panels (a), (b), and (c) is lost entirely; rather, an electron-rich region has formed around the fluorine (at the center of which is an electron-deficient region, as in all other cases). An electron-deficient region has also formed toward the exterior of the compound that is unique to cavity-embedded (NaF)$^{-}$ at this geometry. These pronounced density deformations at the fluorine center could be responsible for the destabilization of (NaF)$^{-}$ at this geometry, relative to the one in which the molecular axis is aligned with the cavity polarization.

\section{Conclusions}

\label{SEC:CONCLUSIONS}

We have developed a computer implementation of the quantum electrodynamics coupled-cluster approach, approximated at the level of coupled single and double electron transitions and single photon transitions (QED-CCSD-1). Following Ref.~\citenum{Koch20_041043}, the complexity of the working equations is minimized via explicit similarity transformation of the Hamiltonian with the singles cluster amplitudes.\cite{Helgaker94_233} We also maximize the efficiency of the $t_1$-transformation through the use of the density-fitting approximation to the two-electron repulsion integral tensor. This implementation of QED-CCSD-1 has been developed within the public version of the \texttt{hilbert} package.\cite{hilbert}

We applied {\color{black}unrestricted} QED-CCSD-1 and QED-HF methods to the description of cavity-induced changes to IPs and EAs in a series of sodium halide compounds (NaX, X = F, Cl, Br, I). In general, IPs were found to be insensitive to cavity interactions, while significant changes to EAs were realized with experimentally-accessible cavity parameters. For a single-mode cavity with a frequency $\omega_{\rm cav}=2.0$ eV and coupling strength $\lambda = 0.05$, the QED-CCSD-1-derived EA for NaF can be reduced from the isolated-compound value of 0.43 eV to only 0.21 eV. Changes predicted by QED-HF are even larger; in the case of NaF, the isolated-compound EA of 0.37 eV is reduced to -0.01 eV for the cavity-coupled system. The substantial deviations between QED-HF and QED-CCSD-1 predictions of the sensitivity of EAs to cavity interactions highlight the importance of electron-electron and electron-photon correlation effects when predicting the properties of cavity-embedded molecules. 
We also demonstrated that the large cavity-induced changes in EAs predicted by QED-HF are associated with considerable  deformations in the electronic density of (NaF)$^{-}$ relative to NaF.  

{\color{black}Aside from the sodium halide compounds discussed in detail in this work, we have also explored cavity-induced changes to EAs and IPs in other classes of sodium compounds. For example, at the unrestricted QED-CCSD-1/def2-TZVPPD level of theory, when interacting with a single-mode cavity (with $\omega_{\rm cav}$ = 2.0 eV and $\lambda$ = 0.05), we have found that the EA of NaOH can be reduced from its cavity-free value of 0.32 eV to only 0.15 eV or 0.10 eV when the cavity mode is aligned with the molecular axis or perpendicular to it, respectively. Similarly, at the same level of theory, the EA of NaCN can be reduced from its cavity-free limit of 0.75 eV to  0.61 eV or 0.55 eV, again, depending on the relative orientation of the cavity mode and molecular axis. As in the case of the sodium halide compounds, the IPs in NaOH and NaCN are less sensitive to cavity interactions, changing by, at most, 0.05 eV. 

We have also explored cavity-induced changes to IPs and EAs in other classes of molecules ({\em i.e.}, organic molecules) and have observed changes to both IPs and EAs of nearly 0.1 eV; however, spin contamination becomes an issue in some of these species, complicating our analysis. }
{\color{black}Recall that the formalism used herein is an unrestricted one and that the resulting QED-HF and QED-CCSD-1 solutions are thus potentially spin contaminated. Fortunately, for the charged sodium compounds, spin contamination does not appear to be an issue. $\langle S^2 \rangle$ values computed at the QED-HF level of theory deviated from the expected values by at most 0.01 ({\em i.e.}, $\langle S^2 \rangle$ = 0.76), and no spin-broken singlet solutions were identified for the neutral compounds. Given the small degree of spin contamination at the reference level, it is reasonable to expect that, for these species, spin contamination at the QED-CCSD-1 level of theory is also not an issue.}

{\color{black}Moving forward, it would be desirable to predict {\em a priori} which ground-state properties, if any, could be modulated through cavity interactions. In the Pauli-Fierez Hamiltonian, in the coherent-state basis, we see that the cavity effects should be largest for molecules with large variances in the dipole moment, with the relevant term in Eq.~\ref{EQN:PFH_COHERENT} being $\frac{1}{2}({\bm \lambda} \cdot[{\bm \mu} - \langle {\bm \mu }\rangle] )^2$. At the self-consistent field (SCF) level, one can easily calculate this quantity for a given isolated molecule in order to estimate its sensitivity to the introduction of the cavity. For the sodium compounds we considered in this work, the anion species have the largest dipole variances (at the SCF level), which is consistent with our observation that the anion is indeed the most sensitive charge state to the cavity.}

\begin{acknowledgments}This material is based upon work supported by the National Science Foundation under Grant No. CHE-1554354.\\
\end{acknowledgments}

\noindent {\bf DATA AVAILABILITY}\\

    The data that support the findings of this study are available from the corresponding author upon reasonable request.


	\bibliography{main}

\begin{thebibliography}{48}%
\makeatletter
\providecommand \@ifxundefined [1]{%
 \@ifx{#1\undefined}
}%
\providecommand \@ifnum [1]{%
 \ifnum #1\expandafter \@firstoftwo
 \else \expandafter \@secondoftwo
 \fi
}%
\providecommand \@ifx [1]{%
 \ifx #1\expandafter \@firstoftwo
 \else \expandafter \@secondoftwo
 \fi
}%
\providecommand \natexlab [1]{#1}%
\providecommand \enquote  [1]{``#1''}%
\providecommand \bibnamefont  [1]{#1}%
\providecommand \bibfnamefont [1]{#1}%
\providecommand \citenamefont [1]{#1}%
\providecommand \href@noop [0]{\@secondoftwo}%
\providecommand \href [0]{\begingroup \@sanitize@url \@href}%
\providecommand \@href[1]{\@@startlink{#1}\@@href}%
\providecommand \@@href[1]{\endgroup#1\@@endlink}%
\providecommand \@sanitize@url [0]{\catcode `\\12\catcode `\$12\catcode
  `\&12\catcode `\#12\catcode `\^12\catcode `\_12\catcode `\%12\relax}%
\providecommand \@@startlink[1]{}%
\providecommand \@@endlink[0]{}%
\providecommand \url  [0]{\begingroup\@sanitize@url \@url }%
\providecommand \@url [1]{\endgroup\@href {#1}{\urlprefix }}%
\providecommand \urlprefix  [0]{URL }%
\providecommand \Eprint [0]{\href }%
\providecommand \doibase [0]{http://dx.doi.org/}%
\providecommand \selectlanguage [0]{\@gobble}%
\providecommand \bibinfo  [0]{\@secondoftwo}%
\providecommand \bibfield  [0]{\@secondoftwo}%
\providecommand \translation [1]{[#1]}%
\providecommand \BibitemOpen [0]{}%
\providecommand \bibitemStop [0]{}%
\providecommand \bibitemNoStop [0]{.\EOS\space}%
\providecommand \EOS [0]{\spacefactor3000\relax}%
\providecommand \BibitemShut  [1]{\csname bibitem#1\endcsname}%
\let\auto@bib@innerbib\@empty
\bibitem [{\citenamefont {Frisk~Kockum}\ \emph {et~al.}(2019)\citenamefont
  {Frisk~Kockum}, \citenamefont {Miranowicz}, \citenamefont {De~Liberato},
  \citenamefont {Savasta},\ and\ \citenamefont {Nori}}]{Nori19_19}%
  \BibitemOpen
  \bibfield  {author} {\bibinfo {author} {\bibfnamefont {A.}~\bibnamefont
  {Frisk~Kockum}}, \bibinfo {author} {\bibfnamefont {A.}~\bibnamefont
  {Miranowicz}}, \bibinfo {author} {\bibfnamefont {S.}~\bibnamefont
  {De~Liberato}}, \bibinfo {author} {\bibfnamefont {S.}~\bibnamefont
  {Savasta}}, \ and\ \bibinfo {author} {\bibfnamefont {F.}~\bibnamefont
  {Nori}},\ }\bibfield  {title} {\enquote {\bibinfo {title} {Ultrastrong
  coupling between light and matter},}\ }\href {\doibase
  10.1038/s42254-018-0006-2} {\bibfield  {journal} {\bibinfo  {journal} {Nature
  Reviews Physics}\ }\textbf {\bibinfo {volume} {1}},\ \bibinfo {pages}
  {19--40} (\bibinfo {year} {2019})}\BibitemShut {NoStop}%
\bibitem [{\citenamefont {Flick}, \citenamefont {Rivera},\ and\ \citenamefont
  {Narang}(2018)}]{Narang18_1479}%
  \BibitemOpen
  \bibfield  {author} {\bibinfo {author} {\bibfnamefont {J.}~\bibnamefont
  {Flick}}, \bibinfo {author} {\bibfnamefont {N.}~\bibnamefont {Rivera}}, \
  and\ \bibinfo {author} {\bibfnamefont {P.}~\bibnamefont {Narang}},\
  }\bibfield  {title} {\enquote {\bibinfo {title} {Strong light-matter coupling
  in quantum chemistry and quantum photonics},}\ }\href {\doibase
  https://doi.org/10.1515/nanoph-2018-0067} {\bibfield  {journal} {\bibinfo
  {journal} {Nanophotonics}\ }\textbf {\bibinfo {volume} {7}},\ \bibinfo
  {pages} {1479 -- 1501} (\bibinfo {year} {2018})}\BibitemShut {NoStop}%
\bibitem [{\citenamefont {Törmä}\ and\ \citenamefont
  {Barnes}(2014)}]{Barnes14_013901}%
  \BibitemOpen
  \bibfield  {author} {\bibinfo {author} {\bibfnamefont {P.}~\bibnamefont
  {Törmä}}\ and\ \bibinfo {author} {\bibfnamefont {W.~L.}\ \bibnamefont
  {Barnes}},\ }\bibfield  {title} {\enquote {\bibinfo {title} {Strong coupling
  between surface plasmon polaritons and emitters: a review},}\ }\href
  {\doibase 10.1088/0034-4885/78/1/013901} {\bibfield  {journal} {\bibinfo
  {journal} {Reports on Progress in Physics}\ }\textbf {\bibinfo {volume}
  {78}},\ \bibinfo {pages} {013901} (\bibinfo {year} {2014})}\BibitemShut
  {NoStop}%
\bibitem [{\citenamefont {Lidzey}\ \emph {et~al.}(1998)\citenamefont {Lidzey},
  \citenamefont {Bradley}, \citenamefont {Skolnick}, \citenamefont {Virgili},
  \citenamefont {Walker},\ and\ \citenamefont {Whittaker}}]{Whittaker98_6697}%
  \BibitemOpen
  \bibfield  {author} {\bibinfo {author} {\bibfnamefont {D.~G.}\ \bibnamefont
  {Lidzey}}, \bibinfo {author} {\bibfnamefont {D.~D.~C.}\ \bibnamefont
  {Bradley}}, \bibinfo {author} {\bibfnamefont {M.~S.}\ \bibnamefont
  {Skolnick}}, \bibinfo {author} {\bibfnamefont {T.}~\bibnamefont {Virgili}},
  \bibinfo {author} {\bibfnamefont {S.}~\bibnamefont {Walker}}, \ and\ \bibinfo
  {author} {\bibfnamefont {D.~M.}\ \bibnamefont {Whittaker}},\ }\bibfield
  {title} {\enquote {\bibinfo {title} {Strong exciton--photon coupling in an
  organic semiconductor microcavity},}\ }\href {\doibase 10.1038/25692}
  {\bibfield  {journal} {\bibinfo  {journal} {Nature}\ }\textbf {\bibinfo
  {volume} {395}},\ \bibinfo {pages} {53--55} (\bibinfo {year}
  {1998})}\BibitemShut {NoStop}%
\bibitem [{\citenamefont {Bellessa}\ \emph {et~al.}(2004)\citenamefont
  {Bellessa}, \citenamefont {Bonnand}, \citenamefont {Plenet},\ and\
  \citenamefont {Mugnier}}]{Mugnier04_036404}%
  \BibitemOpen
  \bibfield  {author} {\bibinfo {author} {\bibfnamefont {J.}~\bibnamefont
  {Bellessa}}, \bibinfo {author} {\bibfnamefont {C.}~\bibnamefont {Bonnand}},
  \bibinfo {author} {\bibfnamefont {J.~C.}\ \bibnamefont {Plenet}}, \ and\
  \bibinfo {author} {\bibfnamefont {J.}~\bibnamefont {Mugnier}},\ }\bibfield
  {title} {\enquote {\bibinfo {title} {Strong coupling between surface plasmons
  and excitons in an organic semiconductor},}\ }\href {\doibase
  10.1103/PhysRevLett.93.036404} {\bibfield  {journal} {\bibinfo  {journal}
  {Physical Review Letters}\ }\textbf {\bibinfo {volume} {93}},\ \bibinfo
  {pages} {036404} (\bibinfo {year} {2004})}\BibitemShut {NoStop}%
\bibitem [{\citenamefont {Hutchison}\ \emph {et~al.}(2012)\citenamefont
  {Hutchison}, \citenamefont {Schwartz}, \citenamefont {Genet}, \citenamefont
  {Devaux},\ and\ \citenamefont {Ebbesen}}]{Ebbesen12_1592}%
  \BibitemOpen
  \bibfield  {author} {\bibinfo {author} {\bibfnamefont {J.~A.}\ \bibnamefont
  {Hutchison}}, \bibinfo {author} {\bibfnamefont {T.}~\bibnamefont {Schwartz}},
  \bibinfo {author} {\bibfnamefont {C.}~\bibnamefont {Genet}}, \bibinfo
  {author} {\bibfnamefont {E.}~\bibnamefont {Devaux}}, \ and\ \bibinfo {author}
  {\bibfnamefont {T.~W.}\ \bibnamefont {Ebbesen}},\ }\bibfield  {title}
  {\enquote {\bibinfo {title} {Modifying chemical landscapes by coupling to
  vacuum fields},}\ }\href {\doibase 10.1002/anie.201107033} {\bibfield
  {journal} {\bibinfo  {journal} {Angewandte Chemie International Edition}\
  }\textbf {\bibinfo {volume} {51}},\ \bibinfo {pages} {1592--1596} (\bibinfo
  {year} {2012})}\BibitemShut {NoStop}%
\bibitem [{\citenamefont {Coles}\ \emph {et~al.}(2014)\citenamefont {Coles},
  \citenamefont {Yang}, \citenamefont {Wang}, \citenamefont {Grant},
  \citenamefont {Taylor}, \citenamefont {Saikin}, \citenamefont {Aspuru-Guzik},
  \citenamefont {Lidzey}, \citenamefont {Tang},\ and\ \citenamefont
  {Smith}}]{Smith14_5561}%
  \BibitemOpen
  \bibfield  {author} {\bibinfo {author} {\bibfnamefont {D.~M.}\ \bibnamefont
  {Coles}}, \bibinfo {author} {\bibfnamefont {Y.}~\bibnamefont {Yang}},
  \bibinfo {author} {\bibfnamefont {Y.}~\bibnamefont {Wang}}, \bibinfo {author}
  {\bibfnamefont {R.~T.}\ \bibnamefont {Grant}}, \bibinfo {author}
  {\bibfnamefont {R.~A.}\ \bibnamefont {Taylor}}, \bibinfo {author}
  {\bibfnamefont {S.~K.}\ \bibnamefont {Saikin}}, \bibinfo {author}
  {\bibfnamefont {A.}~\bibnamefont {Aspuru-Guzik}}, \bibinfo {author}
  {\bibfnamefont {D.~G.}\ \bibnamefont {Lidzey}}, \bibinfo {author}
  {\bibfnamefont {J.~K.-H.}\ \bibnamefont {Tang}}, \ and\ \bibinfo {author}
  {\bibfnamefont {J.~M.}\ \bibnamefont {Smith}},\ }\bibfield  {title} {\enquote
  {\bibinfo {title} {Strong coupling between chlorosomes of photosynthetic
  bacteria and a confined optical cavity mode},}\ }\href {\doibase
  10.1038/ncomms6561} {\bibfield  {journal} {\bibinfo  {journal} {Nature
  Communications}\ }\textbf {\bibinfo {volume} {5}},\ \bibinfo {pages} {5561}
  (\bibinfo {year} {2014})}\BibitemShut {NoStop}%
\bibitem [{\citenamefont {Orgiu}\ \emph {et~al.}(2015)\citenamefont {Orgiu},
  \citenamefont {George}, \citenamefont {Hutchison}, \citenamefont {Devaux},
  \citenamefont {Dayen}, \citenamefont {Doudin}, \citenamefont {Stellacci},
  \citenamefont {Genet}, \citenamefont {Schachenmayer}, \citenamefont {Genes},
  \citenamefont {Pupillo}, \citenamefont {Samor{\`i}},\ and\ \citenamefont
  {Ebbesen}}]{Ebbesen15_1123}%
  \BibitemOpen
  \bibfield  {author} {\bibinfo {author} {\bibfnamefont {E.}~\bibnamefont
  {Orgiu}}, \bibinfo {author} {\bibfnamefont {J.}~\bibnamefont {George}},
  \bibinfo {author} {\bibfnamefont {J.~A.}\ \bibnamefont {Hutchison}}, \bibinfo
  {author} {\bibfnamefont {E.}~\bibnamefont {Devaux}}, \bibinfo {author}
  {\bibfnamefont {J.~F.}\ \bibnamefont {Dayen}}, \bibinfo {author}
  {\bibfnamefont {B.}~\bibnamefont {Doudin}}, \bibinfo {author} {\bibfnamefont
  {F.}~\bibnamefont {Stellacci}}, \bibinfo {author} {\bibfnamefont
  {C.}~\bibnamefont {Genet}}, \bibinfo {author} {\bibfnamefont
  {J.}~\bibnamefont {Schachenmayer}}, \bibinfo {author} {\bibfnamefont
  {C.}~\bibnamefont {Genes}}, \bibinfo {author} {\bibfnamefont
  {G.}~\bibnamefont {Pupillo}}, \bibinfo {author} {\bibfnamefont
  {P.}~\bibnamefont {Samor{\`i}}}, \ and\ \bibinfo {author} {\bibfnamefont
  {T.~W.}\ \bibnamefont {Ebbesen}},\ }\bibfield  {title} {\enquote {\bibinfo
  {title} {Conductivity in organic semiconductors hybridized with the vacuum
  field},}\ }\href {\doibase 10.1038/nmat4392} {\bibfield  {journal} {\bibinfo
  {journal} {Nature Materials}\ }\textbf {\bibinfo {volume} {14}},\ \bibinfo
  {pages} {1123--1129} (\bibinfo {year} {2015})}\BibitemShut {NoStop}%
\bibitem [{\citenamefont {Chikkaraddy}\ \emph {et~al.}(2016)\citenamefont
  {Chikkaraddy}, \citenamefont {de~Nijs}, \citenamefont {Benz}, \citenamefont
  {Barrow}, \citenamefont {Scherman}, \citenamefont {Rosta}, \citenamefont
  {Demetriadou}, \citenamefont {Fox}, \citenamefont {Hess},\ and\ \citenamefont
  {Baumberg}}]{Baumberg16_127}%
  \BibitemOpen
  \bibfield  {author} {\bibinfo {author} {\bibfnamefont {R.}~\bibnamefont
  {Chikkaraddy}}, \bibinfo {author} {\bibfnamefont {B.}~\bibnamefont
  {de~Nijs}}, \bibinfo {author} {\bibfnamefont {F.}~\bibnamefont {Benz}},
  \bibinfo {author} {\bibfnamefont {S.~J.}\ \bibnamefont {Barrow}}, \bibinfo
  {author} {\bibfnamefont {O.~A.}\ \bibnamefont {Scherman}}, \bibinfo {author}
  {\bibfnamefont {E.}~\bibnamefont {Rosta}}, \bibinfo {author} {\bibfnamefont
  {A.}~\bibnamefont {Demetriadou}}, \bibinfo {author} {\bibfnamefont
  {P.}~\bibnamefont {Fox}}, \bibinfo {author} {\bibfnamefont {O.}~\bibnamefont
  {Hess}}, \ and\ \bibinfo {author} {\bibfnamefont {J.~J.}\ \bibnamefont
  {Baumberg}},\ }\bibfield  {title} {\enquote {\bibinfo {title}
  {Single-molecule strong coupling at room temperature in plasmonic
  nanocavities},}\ }\href {\doibase 10.1038/nature17974} {\bibfield  {journal}
  {\bibinfo  {journal} {Nature}\ }\textbf {\bibinfo {volume} {535}},\ \bibinfo
  {pages} {127--130} (\bibinfo {year} {2016})}\BibitemShut {NoStop}%
\bibitem [{\citenamefont {Ebbesen}(2016)}]{Ebbesen16_2403}%
  \BibitemOpen
  \bibfield  {author} {\bibinfo {author} {\bibfnamefont {T.~W.}\ \bibnamefont
  {Ebbesen}},\ }\bibfield  {title} {\enquote {\bibinfo {title} {Hybrid
  light–matter states in a molecular and material science perspective},}\
  }\href {\doibase 10.1021/acs.accounts.6b00295} {\bibfield  {journal}
  {\bibinfo  {journal} {Accounts of Chemical Research}\ }\textbf {\bibinfo
  {volume} {49}},\ \bibinfo {pages} {2403--2412} (\bibinfo {year}
  {2016})}\BibitemShut {NoStop}%
\bibitem [{\citenamefont {Sukharev}\ and\ \citenamefont
  {Nitzan}(2017)}]{Nitzan17_443003}%
  \BibitemOpen
  \bibfield  {author} {\bibinfo {author} {\bibfnamefont {M.}~\bibnamefont
  {Sukharev}}\ and\ \bibinfo {author} {\bibfnamefont {A.}~\bibnamefont
  {Nitzan}},\ }\bibfield  {title} {\enquote {\bibinfo {title} {Optics of
  exciton-plasmon nanomaterials},}\ }\href {\doibase 10.1088/1361-648x/aa85ef}
  {\bibfield  {journal} {\bibinfo  {journal} {Journal of Physics: Condensed
  Matter}\ }\textbf {\bibinfo {volume} {29}},\ \bibinfo {pages} {443003}
  (\bibinfo {year} {2017})}\BibitemShut {NoStop}%
\bibitem [{\citenamefont {Zhong}\ \emph {et~al.}(2017)\citenamefont {Zhong},
  \citenamefont {Chervy}, \citenamefont {Zhang}, \citenamefont {Thomas},
  \citenamefont {George}, \citenamefont {Genet}, \citenamefont {Hutchison},\
  and\ \citenamefont {Ebbesen}}]{Ebbesen17_9034}%
  \BibitemOpen
  \bibfield  {author} {\bibinfo {author} {\bibfnamefont {X.}~\bibnamefont
  {Zhong}}, \bibinfo {author} {\bibfnamefont {T.}~\bibnamefont {Chervy}},
  \bibinfo {author} {\bibfnamefont {L.}~\bibnamefont {Zhang}}, \bibinfo
  {author} {\bibfnamefont {A.}~\bibnamefont {Thomas}}, \bibinfo {author}
  {\bibfnamefont {J.}~\bibnamefont {George}}, \bibinfo {author} {\bibfnamefont
  {C.}~\bibnamefont {Genet}}, \bibinfo {author} {\bibfnamefont {J.~A.}\
  \bibnamefont {Hutchison}}, \ and\ \bibinfo {author} {\bibfnamefont {T.~W.}\
  \bibnamefont {Ebbesen}},\ }\bibfield  {title} {\enquote {\bibinfo {title}
  {Energy transfer between spatially separated entangled molecules},}\ }\href
  {\doibase 10.1002/anie.201703539} {\bibfield  {journal} {\bibinfo  {journal}
  {Angewandte Chemie International Edition}\ }\textbf {\bibinfo {volume}
  {56}},\ \bibinfo {pages} {9034--9038} (\bibinfo {year} {2017})}\BibitemShut
  {NoStop}%
\bibitem [{\citenamefont {Chevrier}\ \emph {et~al.}(2019)\citenamefont
  {Chevrier}, \citenamefont {Benoit}, \citenamefont {Symonds}, \citenamefont
  {Saikin}, \citenamefont {Yuen-Zhou},\ and\ \citenamefont
  {Bellessa}}]{Bellessa19_173902}%
  \BibitemOpen
  \bibfield  {author} {\bibinfo {author} {\bibfnamefont {K.}~\bibnamefont
  {Chevrier}}, \bibinfo {author} {\bibfnamefont {J.~M.}\ \bibnamefont
  {Benoit}}, \bibinfo {author} {\bibfnamefont {C.}~\bibnamefont {Symonds}},
  \bibinfo {author} {\bibfnamefont {S.~K.}\ \bibnamefont {Saikin}}, \bibinfo
  {author} {\bibfnamefont {J.}~\bibnamefont {Yuen-Zhou}}, \ and\ \bibinfo
  {author} {\bibfnamefont {J.}~\bibnamefont {Bellessa}},\ }\bibfield  {title}
  {\enquote {\bibinfo {title} {Anisotropy and controllable band structure in
  suprawavelength polaritonic metasurfaces},}\ }\href {\doibase
  10.1103/PhysRevLett.122.173902} {\bibfield  {journal} {\bibinfo  {journal}
  {Physical Review Letters}\ }\textbf {\bibinfo {volume} {122}},\ \bibinfo
  {pages} {173902} (\bibinfo {year} {2019})}\BibitemShut {NoStop}%
\bibitem [{\citenamefont {K{\'e}na-Cohen}\ and\ \citenamefont
  {Forrest}(2010)}]{Forrest20_371}%
  \BibitemOpen
  \bibfield  {author} {\bibinfo {author} {\bibfnamefont {S.}~\bibnamefont
  {K{\'e}na-Cohen}}\ and\ \bibinfo {author} {\bibfnamefont {S.~R.}\
  \bibnamefont {Forrest}},\ }\bibfield  {title} {\enquote {\bibinfo {title}
  {Room-temperature polariton lasing in an organic single-crystal
  microcavity},}\ }\href {\doibase 10.1038/nphoton.2010.86} {\bibfield
  {journal} {\bibinfo  {journal} {Nature Photonics}\ }\textbf {\bibinfo
  {volume} {4}},\ \bibinfo {pages} {371--375} (\bibinfo {year}
  {2010})}\BibitemShut {NoStop}%
\bibitem [{\citenamefont {Lather}\ \emph {et~al.}(2019)\citenamefont {Lather},
  \citenamefont {Bhatt}, \citenamefont {Thomas}, \citenamefont {Ebbesen},\ and\
  \citenamefont {George}}]{George19_10635}%
  \BibitemOpen
  \bibfield  {author} {\bibinfo {author} {\bibfnamefont {J.}~\bibnamefont
  {Lather}}, \bibinfo {author} {\bibfnamefont {P.}~\bibnamefont {Bhatt}},
  \bibinfo {author} {\bibfnamefont {A.}~\bibnamefont {Thomas}}, \bibinfo
  {author} {\bibfnamefont {T.~W.}\ \bibnamefont {Ebbesen}}, \ and\ \bibinfo
  {author} {\bibfnamefont {J.}~\bibnamefont {George}},\ }\bibfield  {title}
  {\enquote {\bibinfo {title} {Cavity catalysis by cooperative vibrational
  strong coupling of reactant and solvent molecules},}\ }\href {\doibase
  10.1002/anie.201905407} {\bibfield  {journal} {\bibinfo  {journal}
  {Angewandte Chemie International Edition}\ }\textbf {\bibinfo {volume}
  {58}},\ \bibinfo {pages} {10635--10638} (\bibinfo {year} {2019})}\BibitemShut
  {NoStop}%
\bibitem [{\citenamefont {Climent}\ \emph {et~al.}(2019)\citenamefont
  {Climent}, \citenamefont {Galego}, \citenamefont {Garcia-Vidal},\ and\
  \citenamefont {Feist}}]{Feist19_8698}%
  \BibitemOpen
  \bibfield  {author} {\bibinfo {author} {\bibfnamefont {C.}~\bibnamefont
  {Climent}}, \bibinfo {author} {\bibfnamefont {J.}~\bibnamefont {Galego}},
  \bibinfo {author} {\bibfnamefont {F.~J.}\ \bibnamefont {Garcia-Vidal}}, \
  and\ \bibinfo {author} {\bibfnamefont {J.}~\bibnamefont {Feist}},\ }\bibfield
   {title} {\enquote {\bibinfo {title} {Plasmonic nanocavities enable
  self-induced electrostatic catalysis},}\ }\href {\doibase
  10.1002/anie.201901926} {\bibfield  {journal} {\bibinfo  {journal}
  {Angewandte Chemie International Edition}\ }\textbf {\bibinfo {volume}
  {58}},\ \bibinfo {pages} {8698--8702} (\bibinfo {year} {2019})}\BibitemShut
  {NoStop}%
\bibitem [{\citenamefont {Yadav}\ \emph {et~al.}(2020)\citenamefont {Yadav},
  \citenamefont {Otten}, \citenamefont {Wang}, \citenamefont {Cortes},
  \citenamefont {Gosztola}, \citenamefont {Wiederrecht}, \citenamefont {Gray},
  \citenamefont {Odom},\ and\ \citenamefont {Basu}}]{Basu20_5043}%
  \BibitemOpen
  \bibfield  {author} {\bibinfo {author} {\bibfnamefont {R.~K.}\ \bibnamefont
  {Yadav}}, \bibinfo {author} {\bibfnamefont {M.}~\bibnamefont {Otten}},
  \bibinfo {author} {\bibfnamefont {W.}~\bibnamefont {Wang}}, \bibinfo {author}
  {\bibfnamefont {C.~L.}\ \bibnamefont {Cortes}}, \bibinfo {author}
  {\bibfnamefont {D.~J.}\ \bibnamefont {Gosztola}}, \bibinfo {author}
  {\bibfnamefont {G.~P.}\ \bibnamefont {Wiederrecht}}, \bibinfo {author}
  {\bibfnamefont {S.~K.}\ \bibnamefont {Gray}}, \bibinfo {author}
  {\bibfnamefont {T.~W.}\ \bibnamefont {Odom}}, \ and\ \bibinfo {author}
  {\bibfnamefont {J.~K.}\ \bibnamefont {Basu}},\ }\bibfield  {title} {\enquote
  {\bibinfo {title} {Strongly coupled exciton–surface lattice resonances
  engineer long-range energy propagation},}\ }\href {\doibase
  10.1021/acs.nanolett.0c01236} {\bibfield  {journal} {\bibinfo  {journal}
  {Nano Letters}\ }\textbf {\bibinfo {volume} {20}},\ \bibinfo {pages}
  {5043--5049} (\bibinfo {year} {2020})}\BibitemShut {NoStop}%
\bibitem [{\citenamefont {Munkhbat}\ \emph {et~al.}(2018)\citenamefont
  {Munkhbat}, \citenamefont {Wers{\"a}ll}, \citenamefont {Baranov},
  \citenamefont {Antosiewicz},\ and\ \citenamefont
  {Shegai}}]{Shegai18_eaas9552}%
  \BibitemOpen
  \bibfield  {author} {\bibinfo {author} {\bibfnamefont {B.}~\bibnamefont
  {Munkhbat}}, \bibinfo {author} {\bibfnamefont {M.}~\bibnamefont
  {Wers{\"a}ll}}, \bibinfo {author} {\bibfnamefont {D.~G.}\ \bibnamefont
  {Baranov}}, \bibinfo {author} {\bibfnamefont {T.~J.}\ \bibnamefont
  {Antosiewicz}}, \ and\ \bibinfo {author} {\bibfnamefont {T.}~\bibnamefont
  {Shegai}},\ }\bibfield  {title} {\enquote {\bibinfo {title} {Suppression of
  photo-oxidation of organic chromophores by strong coupling to plasmonic
  nanoantennas},}\ }\href {\doibase 10.1126/sciadv.aas9552} {\bibfield
  {journal} {\bibinfo  {journal} {Science Advances}\ }\textbf {\bibinfo
  {volume} {4}},\ \bibinfo {pages} {eaas9552} (\bibinfo {year}
  {2018})}\BibitemShut {NoStop}%
\bibitem [{\citenamefont {Flick}\ \emph {et~al.}(2018)\citenamefont {Flick},
  \citenamefont {Schäfer}, \citenamefont {Ruggenthaler}, \citenamefont
  {Appel},\ and\ \citenamefont {Rubio}}]{Rubio18_992}%
  \BibitemOpen
  \bibfield  {author} {\bibinfo {author} {\bibfnamefont {J.}~\bibnamefont
  {Flick}}, \bibinfo {author} {\bibfnamefont {C.}~\bibnamefont {Schäfer}},
  \bibinfo {author} {\bibfnamefont {M.}~\bibnamefont {Ruggenthaler}}, \bibinfo
  {author} {\bibfnamefont {H.}~\bibnamefont {Appel}}, \ and\ \bibinfo {author}
  {\bibfnamefont {A.}~\bibnamefont {Rubio}},\ }\bibfield  {title} {\enquote
  {\bibinfo {title} {Ab initio optimized effective potentials for real
  molecules in optical cavities: Photon contributions to the molecular ground
  state},}\ }\href {\doibase 10.1021/acsphotonics.7b01279} {\bibfield
  {journal} {\bibinfo  {journal} {ACS Photonics}\ }\textbf {\bibinfo {volume}
  {5}},\ \bibinfo {pages} {992--1005} (\bibinfo {year} {2018})}\BibitemShut
  {NoStop}%
\bibitem [{\citenamefont {Haugland}\ \emph {et~al.}(2020)\citenamefont
  {Haugland}, \citenamefont {Ronca}, \citenamefont {Kj\o{}nstad}, \citenamefont
  {Rubio},\ and\ \citenamefont {Koch}}]{Koch20_041043}%
  \BibitemOpen
  \bibfield  {author} {\bibinfo {author} {\bibfnamefont {T.~S.}\ \bibnamefont
  {Haugland}}, \bibinfo {author} {\bibfnamefont {E.}~\bibnamefont {Ronca}},
  \bibinfo {author} {\bibfnamefont {E.~F.}\ \bibnamefont {Kj\o{}nstad}},
  \bibinfo {author} {\bibfnamefont {A.}~\bibnamefont {Rubio}}, \ and\ \bibinfo
  {author} {\bibfnamefont {H.}~\bibnamefont {Koch}},\ }\bibfield  {title}
  {\enquote {\bibinfo {title} {Coupled cluster theory for molecular polaritons:
  Changing ground and excited states},}\ }\href {\doibase
  10.1103/PhysRevX.10.041043} {\bibfield  {journal} {\bibinfo  {journal}
  {Physical Review X}\ }\textbf {\bibinfo {volume} {10}},\ \bibinfo {pages}
  {041043} (\bibinfo {year} {2020})}\BibitemShut {NoStop}%
\bibitem [{\citenamefont {Ciuti}\ and\ \citenamefont
  {Carusotto}(2006)}]{Carusotto06_033811}%
  \BibitemOpen
  \bibfield  {author} {\bibinfo {author} {\bibfnamefont {C.}~\bibnamefont
  {Ciuti}}\ and\ \bibinfo {author} {\bibfnamefont {I.}~\bibnamefont
  {Carusotto}},\ }\bibfield  {title} {\enquote {\bibinfo {title} {Input-output
  theory of cavities in the ultrastrong coupling regime: The case of
  time-independent cavity parameters},}\ }\href {\doibase
  10.1103/PhysRevA.74.033811} {\bibfield  {journal} {\bibinfo  {journal}
  {Physical Review A}\ }\textbf {\bibinfo {volume} {74}},\ \bibinfo {pages}
  {033811} (\bibinfo {year} {2006})}\BibitemShut {NoStop}%
\bibitem [{\citenamefont {Reitz}, \citenamefont {Sommer},\ and\ \citenamefont
  {Genes}(2019)}]{Genes19_203602}%
  \BibitemOpen
  \bibfield  {author} {\bibinfo {author} {\bibfnamefont {M.}~\bibnamefont
  {Reitz}}, \bibinfo {author} {\bibfnamefont {C.}~\bibnamefont {Sommer}}, \
  and\ \bibinfo {author} {\bibfnamefont {C.}~\bibnamefont {Genes}},\ }\bibfield
   {title} {\enquote {\bibinfo {title} {Langevin approach to quantum optics
  with molecules},}\ }\href {\doibase 10.1103/PhysRevLett.122.203602}
  {\bibfield  {journal} {\bibinfo  {journal} {Physical Review Letters}\
  }\textbf {\bibinfo {volume} {122}},\ \bibinfo {pages} {203602} (\bibinfo
  {year} {2019})}\BibitemShut {NoStop}%
\bibitem [{\citenamefont {Shah}\ \emph {et~al.}(2013)\citenamefont {Shah},
  \citenamefont {Scherer}, \citenamefont {Pelton},\ and\ \citenamefont
  {Gray}}]{Gray13_075411}%
  \BibitemOpen
  \bibfield  {author} {\bibinfo {author} {\bibfnamefont {R.~A.}\ \bibnamefont
  {Shah}}, \bibinfo {author} {\bibfnamefont {N.~F.}\ \bibnamefont {Scherer}},
  \bibinfo {author} {\bibfnamefont {M.}~\bibnamefont {Pelton}}, \ and\ \bibinfo
  {author} {\bibfnamefont {S.~K.}\ \bibnamefont {Gray}},\ }\bibfield  {title}
  {\enquote {\bibinfo {title} {Ultrafast reversal of a fano resonance in a
  plasmon-exciton system},}\ }\href {\doibase 10.1103/PhysRevB.88.075411}
  {\bibfield  {journal} {\bibinfo  {journal} {Physical Review B}\ }\textbf
  {\bibinfo {volume} {88}},\ \bibinfo {pages} {075411} (\bibinfo {year}
  {2013})}\BibitemShut {NoStop}%
\bibitem [{\citenamefont {Mandal}\ and\ \citenamefont
  {Huo}(2019)}]{Huo19_5519}%
  \BibitemOpen
  \bibfield  {author} {\bibinfo {author} {\bibfnamefont {A.}~\bibnamefont
  {Mandal}}\ and\ \bibinfo {author} {\bibfnamefont {P.}~\bibnamefont {Huo}},\
  }\bibfield  {title} {\enquote {\bibinfo {title} {Investigating new
  reactivities enabled by polariton photochemistry},}\ }\href {\doibase
  10.1021/acs.jpclett.9b01599} {\bibfield  {journal} {\bibinfo  {journal} {The
  Journal of Physical Chemistry Letters}\ }\textbf {\bibinfo {volume} {10}},\
  \bibinfo {pages} {5519--5529} (\bibinfo {year} {2019})}\BibitemShut {NoStop}%
\bibitem [{\citenamefont {Mandal}, \citenamefont {Krauss},\ and\ \citenamefont
  {Huo}(2020)}]{Huo20_6321}%
  \BibitemOpen
  \bibfield  {author} {\bibinfo {author} {\bibfnamefont {A.}~\bibnamefont
  {Mandal}}, \bibinfo {author} {\bibfnamefont {T.~D.}\ \bibnamefont {Krauss}},
  \ and\ \bibinfo {author} {\bibfnamefont {P.}~\bibnamefont {Huo}},\ }\bibfield
   {title} {\enquote {\bibinfo {title} {Polariton-mediated electron transfer
  via cavity quantum electrodynamics},}\ }\href {\doibase
  10.1021/acs.jpcb.0c03227} {\bibfield  {journal} {\bibinfo  {journal} {Journal
  of Physical Chemistry B}\ }\textbf {\bibinfo {volume} {124}},\ \bibinfo
  {pages} {6321--6340} (\bibinfo {year} {2020})}\BibitemShut {NoStop}%
\bibitem [{\citenamefont {Antoniou}\ \emph {et~al.}(2020)\citenamefont
  {Antoniou}, \citenamefont {Suchanek}, \citenamefont {Varner},\ and\
  \citenamefont {Foley}}]{Foley20_9063}%
  \BibitemOpen
  \bibfield  {author} {\bibinfo {author} {\bibfnamefont {P.}~\bibnamefont
  {Antoniou}}, \bibinfo {author} {\bibfnamefont {F.}~\bibnamefont {Suchanek}},
  \bibinfo {author} {\bibfnamefont {J.~F.}\ \bibnamefont {Varner}}, \ and\
  \bibinfo {author} {\bibfnamefont {J.~J.}\ \bibnamefont {Foley}},\ }\bibfield
  {title} {\enquote {\bibinfo {title} {Role of cavity losses on nonadiabatic
  couplings and dynamics in polaritonic chemistry},}\ }\href {\doibase
  10.1021/acs.jpclett.0c02406} {\bibfield  {journal} {\bibinfo  {journal}
  {Journal of Physical Chemistry Letters}\ }\textbf {\bibinfo {volume} {11}},\
  \bibinfo {pages} {9063--9069} (\bibinfo {year} {2020})}\BibitemShut {NoStop}%
\bibitem [{\citenamefont {Ruggenthaler}, \citenamefont {Mackenroth},\ and\
  \citenamefont {Bauer}(2011)}]{Bauer11_042107}%
  \BibitemOpen
  \bibfield  {author} {\bibinfo {author} {\bibfnamefont {M.}~\bibnamefont
  {Ruggenthaler}}, \bibinfo {author} {\bibfnamefont {F.}~\bibnamefont
  {Mackenroth}}, \ and\ \bibinfo {author} {\bibfnamefont {D.}~\bibnamefont
  {Bauer}},\ }\bibfield  {title} {\enquote {\bibinfo {title} {Time-dependent
  kohn-sham approach to quantum electrodynamics},}\ }\href {\doibase
  10.1103/PhysRevA.84.042107} {\bibfield  {journal} {\bibinfo  {journal}
  {Physical Review A}\ }\textbf {\bibinfo {volume} {84}},\ \bibinfo {pages}
  {042107} (\bibinfo {year} {2011})}\BibitemShut {NoStop}%
\bibitem [{\citenamefont {Tokatly}(2013)}]{Tokatly13_233001}%
  \BibitemOpen
  \bibfield  {author} {\bibinfo {author} {\bibfnamefont {I.~V.}\ \bibnamefont
  {Tokatly}},\ }\bibfield  {title} {\enquote {\bibinfo {title} {Time-dependent
  density functional theory for many-electron systems interacting with cavity
  photons},}\ }\href {\doibase 10.1103/PhysRevLett.110.233001} {\bibfield
  {journal} {\bibinfo  {journal} {Physical Review Letters}\ }\textbf {\bibinfo
  {volume} {110}},\ \bibinfo {pages} {233001} (\bibinfo {year}
  {2013})}\BibitemShut {NoStop}%
\bibitem [{\citenamefont {Ruggenthaler}\ \emph {et~al.}(2014)\citenamefont
  {Ruggenthaler}, \citenamefont {Flick}, \citenamefont {Pellegrini},
  \citenamefont {Appel}, \citenamefont {Tokatly},\ and\ \citenamefont
  {Rubio}}]{Rubio14_012508}%
  \BibitemOpen
  \bibfield  {author} {\bibinfo {author} {\bibfnamefont {M.}~\bibnamefont
  {Ruggenthaler}}, \bibinfo {author} {\bibfnamefont {J.}~\bibnamefont {Flick}},
  \bibinfo {author} {\bibfnamefont {C.}~\bibnamefont {Pellegrini}}, \bibinfo
  {author} {\bibfnamefont {H.}~\bibnamefont {Appel}}, \bibinfo {author}
  {\bibfnamefont {I.~V.}\ \bibnamefont {Tokatly}}, \ and\ \bibinfo {author}
  {\bibfnamefont {A.}~\bibnamefont {Rubio}},\ }\bibfield  {title} {\enquote
  {\bibinfo {title} {Quantum-electrodynamical density-functional theory:
  Bridging quantum optics and electronic-structure theory},}\ }\href {\doibase
  10.1103/PhysRevA.90.012508} {\bibfield  {journal} {\bibinfo  {journal}
  {Physical Review A}\ }\textbf {\bibinfo {volume} {90}},\ \bibinfo {pages}
  {012508} (\bibinfo {year} {2014})}\BibitemShut {NoStop}%
\bibitem [{\citenamefont {Pellegrini}\ \emph {et~al.}(2015)\citenamefont
  {Pellegrini}, \citenamefont {Flick}, \citenamefont {Tokatly}, \citenamefont
  {Appel},\ and\ \citenamefont {Rubio}}]{Rubio15_093001}%
  \BibitemOpen
  \bibfield  {author} {\bibinfo {author} {\bibfnamefont {C.}~\bibnamefont
  {Pellegrini}}, \bibinfo {author} {\bibfnamefont {J.}~\bibnamefont {Flick}},
  \bibinfo {author} {\bibfnamefont {I.~V.}\ \bibnamefont {Tokatly}}, \bibinfo
  {author} {\bibfnamefont {H.}~\bibnamefont {Appel}}, \ and\ \bibinfo {author}
  {\bibfnamefont {A.}~\bibnamefont {Rubio}},\ }\bibfield  {title} {\enquote
  {\bibinfo {title} {Optimized effective potential for quantum electrodynamical
  time-dependent density functional theory},}\ }\href {\doibase
  10.1103/PhysRevLett.115.093001} {\bibfield  {journal} {\bibinfo  {journal}
  {Physical Review Letters}\ }\textbf {\bibinfo {volume} {115}},\ \bibinfo
  {pages} {093001} (\bibinfo {year} {2015})}\BibitemShut {NoStop}%
\bibitem [{\citenamefont {Jestädt}\ \emph {et~al.}(2019)\citenamefont
  {Jestädt}, \citenamefont {Ruggenthaler}, \citenamefont {Oliveira},
  \citenamefont {Rubio},\ and\ \citenamefont {Appel}}]{Appel19_225}%
  \BibitemOpen
  \bibfield  {author} {\bibinfo {author} {\bibfnamefont {R.}~\bibnamefont
  {Jestädt}}, \bibinfo {author} {\bibfnamefont {M.}~\bibnamefont
  {Ruggenthaler}}, \bibinfo {author} {\bibfnamefont {M.~J.~T.}\ \bibnamefont
  {Oliveira}}, \bibinfo {author} {\bibfnamefont {A.}~\bibnamefont {Rubio}}, \
  and\ \bibinfo {author} {\bibfnamefont {H.}~\bibnamefont {Appel}},\ }\bibfield
   {title} {\enquote {\bibinfo {title} {Light-matter interactions within the
  ehrenfest–maxwell–pauli–kohn–sham framework: fundamentals,
  implementation, and nano-optical applications},}\ }\href {\doibase
  10.1080/00018732.2019.1695875} {\bibfield  {journal} {\bibinfo  {journal}
  {Advances in Physics}\ }\textbf {\bibinfo {volume} {68}},\ \bibinfo {pages}
  {225--333} (\bibinfo {year} {2019})}\BibitemShut {NoStop}%
\bibitem [{\citenamefont {Flick}\ and\ \citenamefont
  {Narang}(2020)}]{Narang20_094116}%
  \BibitemOpen
  \bibfield  {author} {\bibinfo {author} {\bibfnamefont {J.}~\bibnamefont
  {Flick}}\ and\ \bibinfo {author} {\bibfnamefont {P.}~\bibnamefont {Narang}},\
  }\bibfield  {title} {\enquote {\bibinfo {title} {Ab initio polaritonic
  potential-energy surfaces for excited-state nanophotonics and polaritonic
  chemistry},}\ }\href {\doibase 10.1063/5.0021033} {\bibfield  {journal}
  {\bibinfo  {journal} {Journal of Chemical Physics}\ }\textbf {\bibinfo
  {volume} {153}},\ \bibinfo {pages} {094116} (\bibinfo {year}
  {2020})}\BibitemShut {NoStop}%
\bibitem [{\citenamefont {Cohen}, \citenamefont {Mori-S{\'a}nchez},\ and\
  \citenamefont {Yang}(2008)}]{Yang08_792}%
  \BibitemOpen
  \bibfield  {author} {\bibinfo {author} {\bibfnamefont {A.~J.}\ \bibnamefont
  {Cohen}}, \bibinfo {author} {\bibfnamefont {P.}~\bibnamefont
  {Mori-S{\'a}nchez}}, \ and\ \bibinfo {author} {\bibfnamefont
  {W.}~\bibnamefont {Yang}},\ }\bibfield  {title} {\enquote {\bibinfo {title}
  {Insights into current limitations of density functional theory},}\ }\href
  {\doibase 10.1126/science.1158722} {\bibfield  {journal} {\bibinfo  {journal}
  {Science}\ }\textbf {\bibinfo {volume} {321}},\ \bibinfo {pages} {792--794}
  (\bibinfo {year} {2008})}\BibitemShut {NoStop}%
\bibitem [{\citenamefont {{\v{C}}{\'\i}{\v{z}}ek}(1966)}]{Cizek66_4256}%
  \BibitemOpen
  \bibfield  {author} {\bibinfo {author} {\bibfnamefont {J.}~\bibnamefont
  {{\v{C}}{\'\i}{\v{z}}ek}},\ }\bibfield  {title} {\enquote {\bibinfo {title}
  {On the correlation problem in atomic and molecular systems. calculation of
  wavefunction components in ursell‐type expansion using quantum‐field
  theoretical methods},}\ }\href {\doibase 10.1063/1.1727484} {\bibfield
  {journal} {\bibinfo  {journal} {Journal of Chemical Physics}\ }\textbf
  {\bibinfo {volume} {45}},\ \bibinfo {pages} {4256--4266} (\bibinfo {year}
  {1966})}\BibitemShut {NoStop}%
\bibitem [{\citenamefont {{\v{C}}{\'\i}{\v{z}}ek}\ and\ \citenamefont
  {Paldus}(1971)}]{Paldus71_359}%
  \BibitemOpen
  \bibfield  {author} {\bibinfo {author} {\bibfnamefont {J.}~\bibnamefont
  {{\v{C}}{\'\i}{\v{z}}ek}}\ and\ \bibinfo {author} {\bibfnamefont
  {J.}~\bibnamefont {Paldus}},\ }\bibfield  {title} {\enquote {\bibinfo {title}
  {Correlation problems in atomic and molecular systems iii. rederivation of
  the coupled-pair many-electron theory using the traditional quantum chemical
  methods},}\ }\href@noop {} {\bibfield  {journal} {\bibinfo  {journal}
  {International Journal of Quantum Chemistry}\ }\textbf {\bibinfo {volume}
  {5}},\ \bibinfo {pages} {359--379} (\bibinfo {year} {1971})}\BibitemShut
  {NoStop}%
\bibitem [{\citenamefont {Shavitt}\ and\ \citenamefont
  {Bartlett}(2009)}]{Bartlett09_book}%
  \BibitemOpen
  \bibfield  {author} {\bibinfo {author} {\bibfnamefont {I.}~\bibnamefont
  {Shavitt}}\ and\ \bibinfo {author} {\bibfnamefont {R.}~\bibnamefont
  {Bartlett}},\ }\href {https://books.google.com/books?id=SWw6ac1NHZYC} {\emph
  {\bibinfo {title} {Many-Body Methods in Chemistry and Physics: MBPT and
  Coupled-Cluster Theory}}},\ Cambridge Molecular Science\ (\bibinfo
  {publisher} {Cambridge University Press},\ \bibinfo {year}
  {2009})\BibitemShut {NoStop}%
\bibitem [{\citenamefont {Bartlett}\ and\ \citenamefont
  {Musia\l{}}(2007)}]{Musial07_291}%
  \BibitemOpen
  \bibfield  {author} {\bibinfo {author} {\bibfnamefont {R.~J.}\ \bibnamefont
  {Bartlett}}\ and\ \bibinfo {author} {\bibfnamefont {M.}~\bibnamefont
  {Musia\l{}}},\ }\bibfield  {title} {\enquote {\bibinfo {title}
  {Coupled-cluster theory in quantum chemistry},}\ }\href {\doibase
  10.1103/RevModPhys.79.291} {\bibfield  {journal} {\bibinfo  {journal}
  {Reviews of Modern Physics}\ }\textbf {\bibinfo {volume} {79}},\ \bibinfo
  {pages} {291--352} (\bibinfo {year} {2007})}\BibitemShut {NoStop}%
\bibitem [{\citenamefont {Mordovina}\ \emph {et~al.}(2020)\citenamefont
  {Mordovina}, \citenamefont {Bungey}, \citenamefont {Appel}, \citenamefont
  {Knowles}, \citenamefont {Rubio},\ and\ \citenamefont
  {Manby}}]{Manby20_023262}%
  \BibitemOpen
  \bibfield  {author} {\bibinfo {author} {\bibfnamefont {U.}~\bibnamefont
  {Mordovina}}, \bibinfo {author} {\bibfnamefont {C.}~\bibnamefont {Bungey}},
  \bibinfo {author} {\bibfnamefont {H.}~\bibnamefont {Appel}}, \bibinfo
  {author} {\bibfnamefont {P.~J.}\ \bibnamefont {Knowles}}, \bibinfo {author}
  {\bibfnamefont {A.}~\bibnamefont {Rubio}}, \ and\ \bibinfo {author}
  {\bibfnamefont {F.~R.}\ \bibnamefont {Manby}},\ }\bibfield  {title} {\enquote
  {\bibinfo {title} {Polaritonic coupled-cluster theory},}\ }\href {\doibase
  10.1103/PhysRevResearch.2.023262} {\bibfield  {journal} {\bibinfo  {journal}
  {Physical Reviews Research}\ }\textbf {\bibinfo {volume} {2}},\ \bibinfo
  {pages} {023262} (\bibinfo {year} {2020})}\BibitemShut {NoStop}%
\bibitem [{\citenamefont {Spohn}(2004)}]{Spohn04_book}%
  \BibitemOpen
  \bibfield  {author} {\bibinfo {author} {\bibfnamefont {H.}~\bibnamefont
  {Spohn}},\ }\href {https://cds.cern.ch/record/803777} {\emph {\bibinfo
  {title} {{Dynamics of charged particles and their radiation field}}}}\
  (\bibinfo  {publisher} {Cambridge Univ. Press},\ \bibinfo {address}
  {Cambridge},\ \bibinfo {year} {2004})\BibitemShut {NoStop}%
\bibitem [{\citenamefont {Ruggenthaler}\ \emph {et~al.}(2018)\citenamefont
  {Ruggenthaler}, \citenamefont {Tancogne-Dejean}, \citenamefont {Flick},
  \citenamefont {Appel},\ and\ \citenamefont {Rubio}}]{Rubio18_0118}%
  \BibitemOpen
  \bibfield  {author} {\bibinfo {author} {\bibfnamefont {M.}~\bibnamefont
  {Ruggenthaler}}, \bibinfo {author} {\bibfnamefont {N.}~\bibnamefont
  {Tancogne-Dejean}}, \bibinfo {author} {\bibfnamefont {J.}~\bibnamefont
  {Flick}}, \bibinfo {author} {\bibfnamefont {H.}~\bibnamefont {Appel}}, \ and\
  \bibinfo {author} {\bibfnamefont {A.}~\bibnamefont {Rubio}},\ }\bibfield
  {title} {\enquote {\bibinfo {title} {From a quantum-electrodynamical
  light--matter description to novel spectroscopies},}\ }\href {\doibase
  10.1038/s41570-018-0118} {\bibfield  {journal} {\bibinfo  {journal} {Nature
  Reviews Chemistry}\ }\textbf {\bibinfo {volume} {2}},\ \bibinfo {pages}
  {0118} (\bibinfo {year} {2018})}\BibitemShut {NoStop}%
\bibitem [{\citenamefont {Flick}\ \emph {et~al.}(2017)\citenamefont {Flick},
  \citenamefont {Appel}, \citenamefont {Ruggenthaler},\ and\ \citenamefont
  {Rubio}}]{Rubio17_1616}%
  \BibitemOpen
  \bibfield  {author} {\bibinfo {author} {\bibfnamefont {J.}~\bibnamefont
  {Flick}}, \bibinfo {author} {\bibfnamefont {H.}~\bibnamefont {Appel}},
  \bibinfo {author} {\bibfnamefont {M.}~\bibnamefont {Ruggenthaler}}, \ and\
  \bibinfo {author} {\bibfnamefont {A.}~\bibnamefont {Rubio}},\ }\bibfield
  {title} {\enquote {\bibinfo {title} {Cavity born–oppenheimer approximation
  for correlated electron–nuclear-photon systems},}\ }\href {\doibase
  10.1021/acs.jctc.6b01126} {\bibfield  {journal} {\bibinfo  {journal} {Journal
  of Chemical Theory and Computation}\ }\textbf {\bibinfo {volume} {13}},\
  \bibinfo {pages} {1616--1625} (\bibinfo {year} {2017})}\BibitemShut {NoStop}%
\bibitem [{\citenamefont {Nascimento}\ and\ \citenamefont {{DePrince
  III}}(2015)}]{DePrince15_214104}%
  \BibitemOpen
  \bibfield  {author} {\bibinfo {author} {\bibfnamefont {D.~R.}\ \bibnamefont
  {Nascimento}}\ and\ \bibinfo {author} {\bibfnamefont {A.~E.}\ \bibnamefont
  {{DePrince III}}},\ }\bibfield  {title} {\enquote {\bibinfo {title} {Modeling
  molecule-plasmon interactions using quantized radiation fields within
  time-dependent electronic structure theory},}\ }\href {\doibase
  10.1063/1.4936348} {\bibfield  {journal} {\bibinfo  {journal} {Journal of
  Chemical Physics}\ }\textbf {\bibinfo {volume} {143}},\ \bibinfo {pages}
  {214104} (\bibinfo {year} {2015})}\BibitemShut {NoStop}%
\bibitem [{\citenamefont {{DePrince III}}(2020)}]{hilbert}%
  \BibitemOpen
  \bibfield  {author} {\bibinfo {author} {\bibfnamefont {A.~E.}\ \bibnamefont
  {{DePrince III}}},\ }\href@noop {} {\enquote {\bibinfo {title} {{Hilbert: a
  space for quantum chemistry plugins to Psi4}},}\ } (\bibinfo {year} {2020}),\
  \bibinfo {note} {https://github.com/edeprince3/hilbert (last accessed
  November, 2020)}\BibitemShut {NoStop}%
\bibitem [{\citenamefont {Smith}\ \emph {et~al.}(2020)\citenamefont {Smith},
  \citenamefont {Burns}, \citenamefont {Simmonett}, \citenamefont {Parrish},
  \citenamefont {Schieber}, \citenamefont {Galvelis}, \citenamefont {Kraus},
  \citenamefont {Kruse}, \citenamefont {Di~Remigio}, \citenamefont {Alenaizan},
  \citenamefont {James}, \citenamefont {Lehtola}, \citenamefont {Misiewicz},
  \citenamefont {Scheurer}, \citenamefont {Shaw}, \citenamefont {Schriber},
  \citenamefont {Xie}, \citenamefont {Glick}, \citenamefont {Sirianni},
  \citenamefont {O'Brien}, \citenamefont {Waldrop}, \citenamefont {Kumar},
  \citenamefont {Hohenstein}, \citenamefont {Pritchard}, \citenamefont
  {Brooks}, \citenamefont {Schaefer}, \citenamefont {Sokolov}, \citenamefont
  {Patkowski}, \citenamefont {DePrince}, \citenamefont {Bozkaya}, \citenamefont
  {King}, \citenamefont {Evangelista}, \citenamefont {Turney}, \citenamefont
  {Crawford},\ and\ \citenamefont {Sherrill}}]{Sherrill20_184108}%
  \BibitemOpen
  \bibfield  {author} {\bibinfo {author} {\bibfnamefont {D.~G.~A.}\
  \bibnamefont {Smith}}, \bibinfo {author} {\bibfnamefont {L.~A.}\ \bibnamefont
  {Burns}}, \bibinfo {author} {\bibfnamefont {A.~C.}\ \bibnamefont
  {Simmonett}}, \bibinfo {author} {\bibfnamefont {R.~M.}\ \bibnamefont
  {Parrish}}, \bibinfo {author} {\bibfnamefont {M.~C.}\ \bibnamefont
  {Schieber}}, \bibinfo {author} {\bibfnamefont {R.}~\bibnamefont {Galvelis}},
  \bibinfo {author} {\bibfnamefont {P.}~\bibnamefont {Kraus}}, \bibinfo
  {author} {\bibfnamefont {H.}~\bibnamefont {Kruse}}, \bibinfo {author}
  {\bibfnamefont {R.}~\bibnamefont {Di~Remigio}}, \bibinfo {author}
  {\bibfnamefont {A.}~\bibnamefont {Alenaizan}}, \bibinfo {author}
  {\bibfnamefont {A.~M.}\ \bibnamefont {James}}, \bibinfo {author}
  {\bibfnamefont {S.}~\bibnamefont {Lehtola}}, \bibinfo {author} {\bibfnamefont
  {J.~P.}\ \bibnamefont {Misiewicz}}, \bibinfo {author} {\bibfnamefont
  {M.}~\bibnamefont {Scheurer}}, \bibinfo {author} {\bibfnamefont {R.~A.}\
  \bibnamefont {Shaw}}, \bibinfo {author} {\bibfnamefont {J.~B.}\ \bibnamefont
  {Schriber}}, \bibinfo {author} {\bibfnamefont {Y.}~\bibnamefont {Xie}},
  \bibinfo {author} {\bibfnamefont {Z.~L.}\ \bibnamefont {Glick}}, \bibinfo
  {author} {\bibfnamefont {D.~A.}\ \bibnamefont {Sirianni}}, \bibinfo {author}
  {\bibfnamefont {J.~S.}\ \bibnamefont {O'Brien}}, \bibinfo {author}
  {\bibfnamefont {J.~M.}\ \bibnamefont {Waldrop}}, \bibinfo {author}
  {\bibfnamefont {A.}~\bibnamefont {Kumar}}, \bibinfo {author} {\bibfnamefont
  {E.~G.}\ \bibnamefont {Hohenstein}}, \bibinfo {author} {\bibfnamefont
  {B.~P.}\ \bibnamefont {Pritchard}}, \bibinfo {author} {\bibfnamefont {B.~R.}\
  \bibnamefont {Brooks}}, \bibinfo {author} {\bibfnamefont {H.~F.}\
  \bibnamefont {Schaefer}}, \bibinfo {author} {\bibfnamefont {A.~Y.}\
  \bibnamefont {Sokolov}}, \bibinfo {author} {\bibfnamefont {K.}~\bibnamefont
  {Patkowski}}, \bibinfo {author} {\bibfnamefont {A.~E.}\ \bibnamefont
  {DePrince}}, \bibinfo {author} {\bibfnamefont {U.}~\bibnamefont {Bozkaya}},
  \bibinfo {author} {\bibfnamefont {R.~A.}\ \bibnamefont {King}}, \bibinfo
  {author} {\bibfnamefont {F.~A.}\ \bibnamefont {Evangelista}}, \bibinfo
  {author} {\bibfnamefont {J.~M.}\ \bibnamefont {Turney}}, \bibinfo {author}
  {\bibfnamefont {T.~D.}\ \bibnamefont {Crawford}}, \ and\ \bibinfo {author}
  {\bibfnamefont {C.~D.}\ \bibnamefont {Sherrill}},\ }\bibfield  {title}
  {\enquote {\bibinfo {title} {Psi4 1.4: Open-source software for
  high-throughput quantum chemistry},}\ }\href {\doibase 10.1063/5.0006002}
  {\bibfield  {journal} {\bibinfo  {journal} {Journal of Chemical Physics}\
  }\textbf {\bibinfo {volume} {152}},\ \bibinfo {pages} {184108} (\bibinfo
  {year} {2020})}\BibitemShut {NoStop}%
\bibitem [{\citenamefont {Whitten}(1973)}]{Whitten73_4496}%
  \BibitemOpen
  \bibfield  {author} {\bibinfo {author} {\bibfnamefont {J.~L.}\ \bibnamefont
  {Whitten}},\ }\bibfield  {title} {\enquote {\bibinfo {title} {Coulombic
  potential energy integrals and approximations},}\ }\href@noop {} {\bibfield
  {journal} {\bibinfo  {journal} {Journal of Chemical Physics}\ }\textbf
  {\bibinfo {volume} {58}},\ \bibinfo {pages} {4496--4501} (\bibinfo {year}
  {1973})}\BibitemShut {NoStop}%
\bibitem [{\citenamefont {Dunlap}, \citenamefont {Connolly},\ and\
  \citenamefont {Sabin}(1979)}]{Sabin79_3396}%
  \BibitemOpen
  \bibfield  {author} {\bibinfo {author} {\bibfnamefont {B.~I.}\ \bibnamefont
  {Dunlap}}, \bibinfo {author} {\bibfnamefont {J.~W.~D.}\ \bibnamefont
  {Connolly}}, \ and\ \bibinfo {author} {\bibfnamefont {J.~R.}\ \bibnamefont
  {Sabin}},\ }\bibfield  {title} {\enquote {\bibinfo {title} {On some
  approximations in applications of $x_\alpha$ theory},}\ }\href@noop {}
  {\bibfield  {journal} {\bibinfo  {journal} {Journal of Chemical Physics}\
  }\textbf {\bibinfo {volume} {71}},\ \bibinfo {pages} {3396--3402} (\bibinfo
  {year} {1979})}\BibitemShut {NoStop}%
\bibitem [{\citenamefont {Koch}\ \emph {et~al.}(1994)\citenamefont {Koch},
  \citenamefont {Christiansen}, \citenamefont {Kobayashi}, \citenamefont
  {Jørgensen},\ and\ \citenamefont {Helgaker}}]{Helgaker94_233}%
  \BibitemOpen
  \bibfield  {author} {\bibinfo {author} {\bibfnamefont {H.}~\bibnamefont
  {Koch}}, \bibinfo {author} {\bibfnamefont {O.}~\bibnamefont {Christiansen}},
  \bibinfo {author} {\bibfnamefont {R.}~\bibnamefont {Kobayashi}}, \bibinfo
  {author} {\bibfnamefont {P.}~\bibnamefont {Jørgensen}}, \ and\ \bibinfo
  {author} {\bibfnamefont {T.}~\bibnamefont {Helgaker}},\ }\bibfield  {title}
  {\enquote {\bibinfo {title} {A direct atomic orbital driven implementation of
  the coupled cluster singles and doubles (ccsd) model},}\ }\href {\doibase
  https://doi.org/10.1016/0009-2614(94)00898-1} {\bibfield  {journal} {\bibinfo
   {journal} {Chemical Physics Letters}\ }\textbf {\bibinfo {volume} {228}},\
  \bibinfo {pages} {233 -- 238} (\bibinfo {year} {1994})}\BibitemShut {NoStop}%
\bibitem [{\citenamefont {{DePrince III}}\ and\ \citenamefont
  {Sherrill}(2013)}]{Sherrill13_2687}%
  \BibitemOpen
  \bibfield  {author} {\bibinfo {author} {\bibfnamefont {A.~E.}\ \bibnamefont
  {{DePrince III}}}\ and\ \bibinfo {author} {\bibfnamefont {C.~D.}\
  \bibnamefont {Sherrill}},\ }\bibfield  {title} {\enquote {\bibinfo {title}
  {Accuracy and efficiency of coupled-cluster theory using density
  fitting/cholesky decomposition, frozen natural orbitals, and a t1-transformed
  hamiltonian},}\ }\href {\doibase 10.1021/ct400250u} {\bibfield  {journal}
  {\bibinfo  {journal} {Journal of Chemical Theory and Computation}\ }\textbf
  {\bibinfo {volume} {9}},\ \bibinfo {pages} {2687--2696} (\bibinfo {year}
  {2013})}\BibitemShut {NoStop}%
\end{thebibliography}%

\end{document}